\documentclass[sn-mathphys]{sn-jnl}

\usepackage{subfig} 
\usepackage{booktabs}
\usepackage{amsmath}
\usepackage{xcolor}
\usepackage{adjustbox}
\usepackage{multicol}
\usepackage{enumitem}
\usepackage{xcolor}
\usepackage{float}
\definecolor{MyBlue}{rgb}{0.25,0.5,0.75}
\colorlet{NextBlue}{MyBlue!20}
\colorlet{SecondBlue}{MyBlue!40}
\jyear{2021}

\begin{document}

\title[Efficient Simulation of the Heat Transfer in Fused Filament Fabrication]{Efficient Simulation of the Heat Transfer in Fused Filament Fabrication}

\author*[1,2]{\fnm{Nathalie} \sur{Ramos}}\email{nathalie.ramos@ntnu.no}
\author[2]{\fnm{Christoph} \sur{Mittermeier}}
\author[2,1]{\fnm{Josef} \sur{Kiendl}}

\affil[1]{\orgdiv{Department of Marine Technology}, \orgname{Norwegian University of Science and Technology}, \orgaddress{\street{Otto Nielsens Veg 10}, \city{Trondheim}, \postcode{7491}, \country{Norway}}}

\affil[2]{\orgdiv{Institute of Engineering Mechanics \& Structural Mechanics}, \orgname{Bundeswehr University Munich}, \orgaddress{\street{Werner-Heisenberg-Weg 39}, \city{Neubiberg}, \postcode{85577}, \country{Germany}}}

\abstract{Heat transfer simulations of the fused filament fabrication process are an important tool to predict bonding, residual stresses and strength of 3D printed parts. But in order to capture the significant thermal gradients that occur in the FFF printing process, a fine mesh discretization and short time steps are required, leading to extensive computational efforts. In this work a simulation framework is presented which combines several efficiency measures with the objective of reducing the computational efforts required in simulating the FFF printing process without simplifying the deposition physics or reducing the overall accuracy. Thus, the material deposition has been modeled with a hybrid element activation approach and elements are adaptively coarsened through an error-based coarsening condition. Additionally, an appropriate coarsening technique is presented for geometries with air-filled infill patterns. The accuracy of the numerical framework is experimentally validated and the efficiency of the framework is validated numerically by comparing the performance of models with and without any efficiency measures. Finally, its effectiveness is shown by simulating the printing process of a larger geometry.}


\maketitle

\section{Introduction}
\label{sec:intro}

Fused filament fabrication (FFF) is one of the additive manufacturing (AM) or three-dimensional (3D) printing technologies in which hot polymer is extruded in a layer-by-layer fashion along a predetermined path to form a 3D object. FFF is the most commonly-used AM technique due to advantages such as the wide availability of low-price materials, its easy operability and low energy requirements \cite{Fitzharris2018},\cite{Gao2021},\cite{Zhang2018}. Although FFF is moving from being primarily a prototyping tool into being a manufacturing tool, the mechanical anisotropy and low mechanical strength of FFF printed parts in comparison to parts produced with traditional polymer manufacturing methods hinder this evolution from happening \cite{Gao2021},\cite{Wong2012},\cite{Vanaei2021}. One of the causes is the discontinuous nature of the FFF process: a molten fiber is extruded and deposited onto the previously deposited layer, forming bonds with adjacent fibers \cite{Yin2018}. This bond interface between layers tends to be the weakest link in FFF printed parts and the strength in z-direction tends to be much lower than in other directions \cite{Gao2021}. Additionally, the rapid heating and cooling which occurs during the deposition process leads to high thermal gradients which can result in residual stresses \cite{Hajializadeh2019}. This can also impact mechanical strength. In both cases, understanding the heat transfer and its significant impact on the bonding and strength of 3D printed parts is crucial in advancing the applicability of FFF \cite{Vanaei2021}, \cite{Zhang2017}.\\ 
\indent There are several works which address the heat transfer during the FFF printing process, albeit experimentally \cite{Yin2018}, \cite{Xu2020}, \cite{Ferraris2019}, analytically \cite{Costa2015} or numerically. Amico and Peterson used finite element (FE) analysis in COMSOL multiphysics to simulate the deposition of a wall of one road thick \cite{Amico2018}. Simulation of nozzle movement and material deposition was achieved using COMSOL's 'deformed geometry' node. Xu et al. also simulated the heat transfer in the FFF printing of a thin wall by using a 3D FE model implemented in C++ originally developed to study heat exchange during metal selective laser melting \cite{Xu2020}. However, the deposition of polymer fractions was performed by changing properties of the FE domain from air to polymer. Zhou et al. used the element birth and death feature in FE software Ansys to simulate the deposition process of a cuboid shaped thin walled structure \cite{Zhou2017}. Cattenone et al. also used sequential element activation for the simulation of a spring and bridge in Abaqus \cite{Cattenone2019}. An extensive review of further appropriate finite element methods for such heat transfer simulations can be found in \cite{Meyghani2017}.\\
\indent In order to accurately capture the significant thermal gradients in these simulations, a very fine mesh discretization and short time steps are required and in return significant computational effort is required and high physical memory demands must be met \cite{Cattenone2019},\cite{Denlinger2014}. Thus, simulations in many of the aforementioned numerical methods have been performed on a small scale. Several strategies have been proposed to achieve compationally efficient frameworks to simulate AM processes. Dimensional reduction has been used in case of simple geometries which allow for a 2D simplification of the 3D geometry \cite{Umer2019}. Spatial reduction is another method to reduce the computational effort. A hybrid element activation strategy has been employed in various works as a global remeshing approach \cite{Denlinger2014},\cite{Janayath2018},\cite{Michaleris2014}. Instead of having all elements representing the final geometry present from the start of the analysis in an inactive state, the final geometry is discretized in a number of subsequent meshes each to which a new quiet layer has been added. Lastly, adaptive meshing is also often used in a bid to minimize the computational expense. Adaptive refinement is typically implemented to achieve a refined, localised mesh in the vicinity of the melt zone or heat affected zone of the heat source \cite{Zeng2014}, \cite{Baiges2021}. Conversely, entire layers further removed from the heat source can also be adaptively coarsened by lumping them together whilst keeping a homogeneous fine mesh around the heat source. This method has its origins in the numerical simulation of welding and it is now often applied in the simulation of metal AM \cite{Denlinger2014},\cite{Malmelov2019}.\\ 
\indent These aforementioned techniques are mostly applied in simulations of metal printing where geometries generally tend to be larger and a loss of accuracy is often inevitable as a result. In this work a simulation framework combining various techniques is presented which contributes towards achieving a reduced computational effort in thermal simulations of the FFF printing process, without sacrificing their overall accuracy. An adaptive coarsening framework is developed in which elements are gradually coarsened over the height of a printed part when satisfying an error-based condition instead of coarsening at pre-defined moments which can lead to premature coarsening and an increased loss of accuracy. Additionally, remeshing is applied in the form of a hybrid element activation approach to further reduce the number of degrees of freedom present in the finite element meshes and an appropriate coarsening technique is presented for geometries with air-filled infill patterns. This entire framework is presented in detail together with the governing equations describing the transient heat transfer analysis. The accuracy of the numerical framework is experimentally validated and the efficiency of the framework is validated numerically by comparing the performance of models with and without any efficiency measures. Finally, the effectiveness of the simulation framework is tested on a larger geometry.

\section{Experimental Set-Up}
\label{sec:exp}
\begin{figure}
\centering
  \begin{tabular}{c}
    \subfloat[Set-up]{\includegraphics[width=0.4\linewidth]{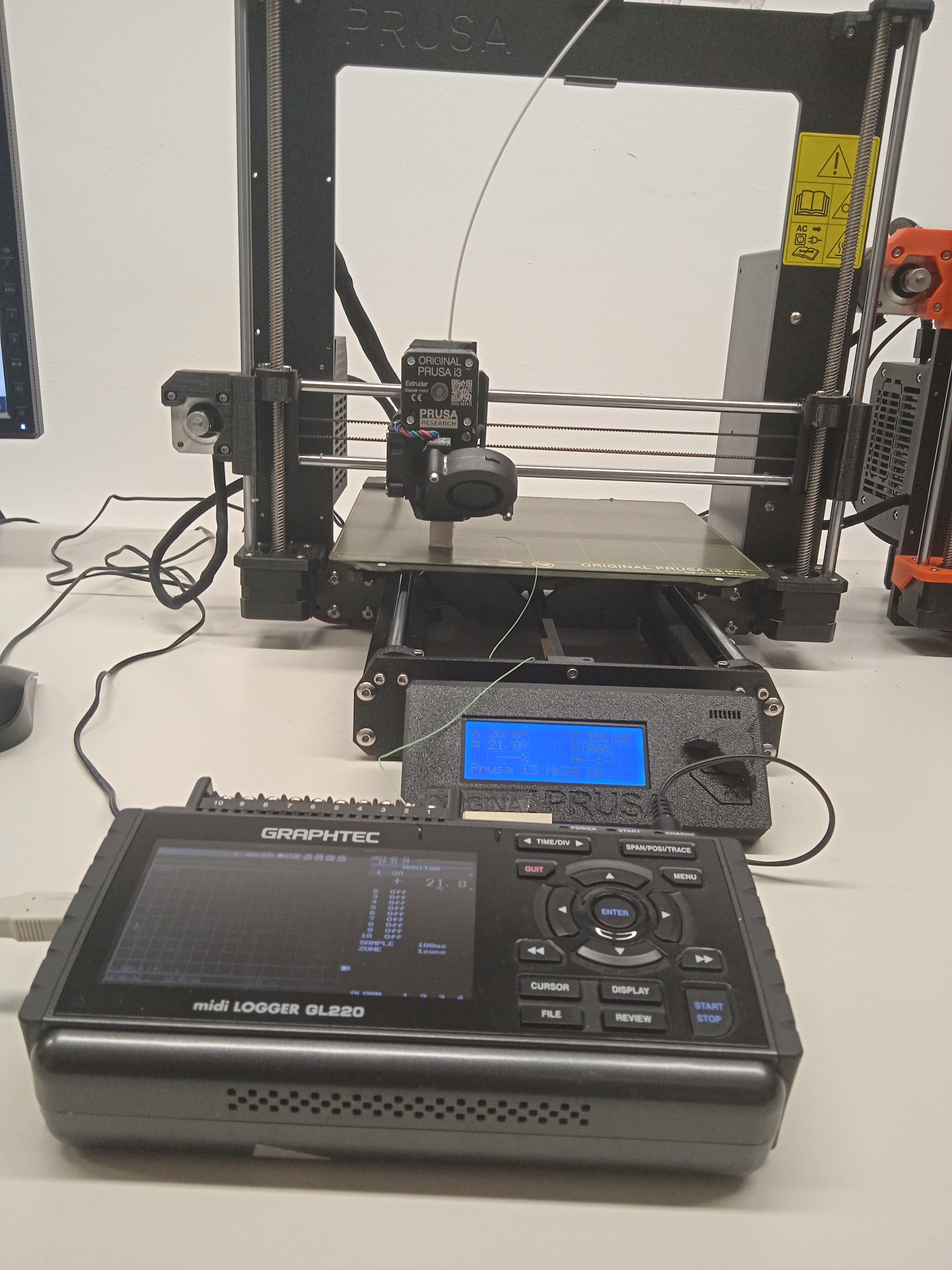}}
  \end{tabular}\hspace{1em}
  \begin{tabular}[m]{c}
    \subfloat[Block geometry with N1, N2, N3]{\includegraphics[width=0.3\linewidth]{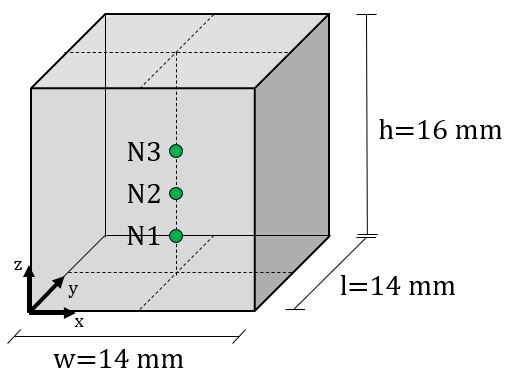}}\\
	\subfloat[Measuring device]{\includegraphics[width=0.3\linewidth]{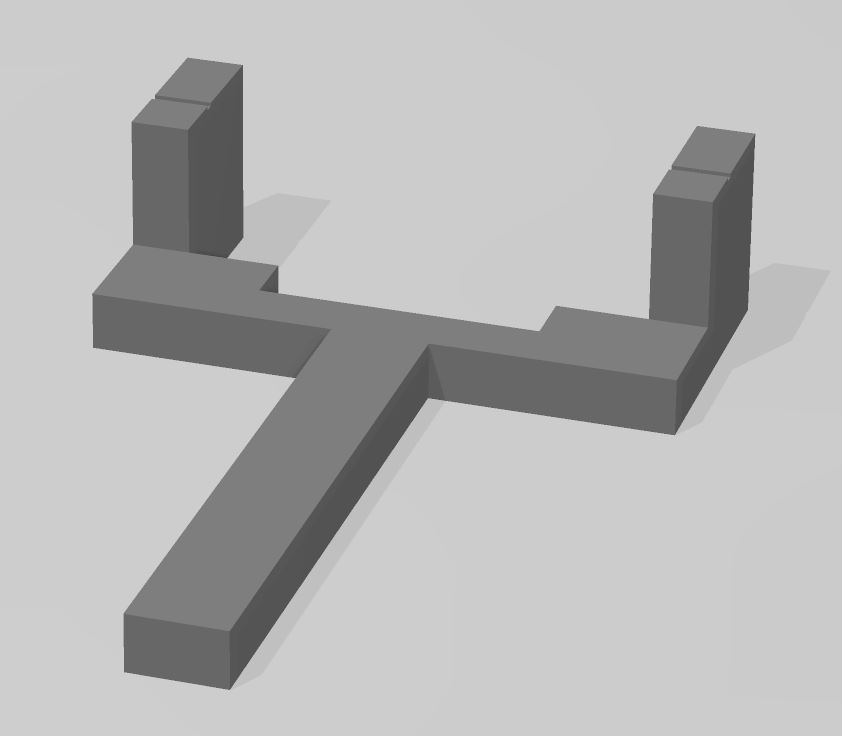}}
  \end{tabular} 
  \caption{Experimental model and set-up}
\label{fig:exp_setup}
\end{figure}

In order to validate the numerical simulations presented in this work, thermal measurements were performed during the printing of a block geometry as shown in figure \ref{fig:exp_setup}. All of the samples were printed with a Prusa i3 MK3 printer. The material used was polylactic acid (PLA) and its thermal properties as provided by the manufacturer Fillamentum are listed in table \ref{tab:mat_prop}. The geometry was printed with the process parameters listed in table \ref{tab:proc_param}. The experimental set-up is shown in figure \ref{fig:exp_setup}. K-type thermocouples were used to measure the temperature during the printing of the block at three measuring locations. N1 was located at z=$\frac{1}{10}\cdot$h, N2 was located at z=$\frac{2}{5}\cdot$h and N3 was located at z=$\frac{3}{5}\cdot$h, where h is the height of the block. All three points were situated in the same vertical planes; x=$\frac{4}{7}\cdot$w and y=$\frac{4}{7}\cdot$l where w and l are the width and length of the block respectively. The temperature was recorded with a Graphtec GL220 data logger with a sampling frequency of 10 Hz. The exact measuring procedure was as follows: the block was initially printed up to the layer where the measuring point was located. The print was then paused for ten seconds to insert the thermocouple at the correct measuring location. This was done by spanning the wire over a printed measuring device (figure \ref{fig:exp_setup}) which controlled the height and depth at which the thermocouple was placed with respect to the block. After placing the thermocouple, printing was resumed and the recording was started. This process was repeated four times for each measuring location. Thus, a total of 12 samples were printed and subjected to temperature recordings.

\begin{table} 
\centering
\begin{tabular}{lccc}
\toprule 
\textbf{Property} & \textbf{PLA} & \textbf{Air}\\
\midrule
Density $\rho$ [kg/m$^{3}$] & 1240 & 1.41 \\
Specific heat capacity $c_p$ [J/kg$\cdot$K] & 1800 & 716  \\
Conductivity $K_{0}$ [W/m$\cdot$K] & 0.13 & 0.023 \\
\bottomrule 
\end{tabular}
\caption{Thermal properties PLA and air}
\label{tab:mat_prop}
\end{table} 

\begin{table} 
\centering
\begin{tabular}{lll}
\toprule 
\textbf{Process parameter} & \textbf{Symbol} & \textbf{Value}\\
\midrule
Printing speed & $v_p$ & 30 mm/s \\
Layer height & $dh$ & 0.2 mm \\
Filament width & $wf$ & 0.5 mm\\
Nozzle temperature & $T_n$ & 210 $^\circ$C\\
Ambient temperature & $T_a$ & 25 $^\circ$C\\
Bed temperature & $T_b$ & 60 $^\circ$C\\
Heat transfer coefficient & $h$ & 25 W/m$^2$ K\\
\bottomrule 
\end{tabular}
\caption{Process parameters FFF}
\label{tab:proc_param}
\end{table}

\section{Numerical methods}
\label{sec:num_met}
The heat transfer that occurs during the FFF printing process presented in section \ref{sec:exp} was numerically simulated by performing a transient thermal analysis. In this section, the finite element model and the efficiency measures to reduce the computional efforts in simulating the transient thermal analysis of FFF are presented.

\subsection{Heat Transfer Analysis}
\label{sec:heat_eq}

Various modes of heat exchange occur during the thermally driven deposition process in FFF \cite{Vanaei2021}. Since the focus in this work is on the heat transfer during the deposition process, all heat transfer mechanisms that occur within the nozzle before and during extrusion are beyond the scope of this work.\\
\indent Starting from the energy balance, the transient heat transfer can be described by the following partial differential equation (PDE) \cite{Habermann2013}:

\begin{equation}
\rho{c}_{p}\frac{\partial T(\textbf{x},t)}{\partial t}=\nabla\cdot(K_{0}\nabla T(\textbf{x},t))+Q
\label{eq:heateq_general}
\end{equation}

\noindent in which $T$ [K] is the temperature, $\rho$ [kg/m$^{3}$] is the material density, $c_p$ [J/kg K] is the specific heat capacity, $K_0$ [W/m K] is the conductivity  and $Q$ [W/m$^{3}$] is the heat source. The left hand side represents the change in the thermal energy storage whereas the first term on the right hand side represents the heat transfer by conduction. The heat flux vector can be identified as 

\begin{equation}
\textbf{q}=-K_0\nabla T(\textbf{x},t)
\label{eq:heat_flux}
\end{equation}

\noindent  Solving the initial value problem given by the PDE in eq. \ref{eq:heateq_general} requires specification of the initial conditions at every point in the considered domain and specification of the temperatures along the boundary (Dirichlet boundary conditions) or its derivatives (Neumann boundary conditions). The heat flow due to convection is given by Newton's law of cooling which states that the heat energy flowing out per unit time per unit surface is proportional to the difference between the surface temperature $T_s$ [K] and the temperature outside the surface $T_{\infty}$ [K]:
  
\begin{equation}
-K_0\nabla T(\textbf{x},t)\mathbf{\cdot n}=h(T_s-T_{\infty})
\label{eq:bc_conv}
\end{equation}

\noindent where $\mathbf{n}$ is the unit vector that points to the outer normal and where $h$ [W/m$^{2}$ K] is the convective heat transfer coefficient. Finally, the heat flux due to radiation is defined by the Stefan-Boltzmann law:

\begin{equation}
-K_0\nabla T(\textbf{x},t)\mathbf{\cdot n}=\varepsilon\sigma(T_{s}^{4}-T_{\infty}^{4})
\label{eq:bc_rad}
\end{equation}

\noindent where $\varepsilon$ is the emissivity and $\sigma$ is the Stefan-Boltzmann constant.  

\subsection{Modeling the Material Deposition}
\label{subsec:mat_dep}
The continuous material deposition on the build stage or on previously deposited layers was simulated by using sequential element activation. In such an analysis all finite elements representing the fully printed geometry are discretized in the finite element mesh and they are deactivated at the start of the analysis. The deposition process is then simulated by sequentially activating elements in the subsequent time steps along the path of the printing nozzle until the full geometry is activated. An element is initially deactivated by reducing the conductivity and specific heat capacity to near-zero values. Thus, the elements are still present in the FE mesh and the attached degrees of freedom (dofs) are present in the global system of equations, but they don't influence the solution. Conversely, the material properties are restored to their original values upon element activation. \\
\indent The numerical stability and accuracy of such an analysis will depend on the time step size. The time step size was determined by calculating the time required to activate or deposit a single element:

\begin{equation}
\Delta t=\frac{dl}{v_p}
\end{equation}
\noindent where $dl$ is the dimension of the element in the traveling direction of the nozzle, and $v_p$ is the printing speed. The objective of this work was not to determine an optimal time step size as there are other works which have dedicated significant efforts to this topic \cite{Cattenone2019}. $\Delta t$ used in the current paper was significantly smaller that what is necessary to capture the cooling rate of PLA, so it was assumed to be small enough.\\

\subsection{Loads and Boundary Conditions}
\label{subsec:loads_bcs}
\begin{figure}
\centering
\includegraphics[scale=0.3]{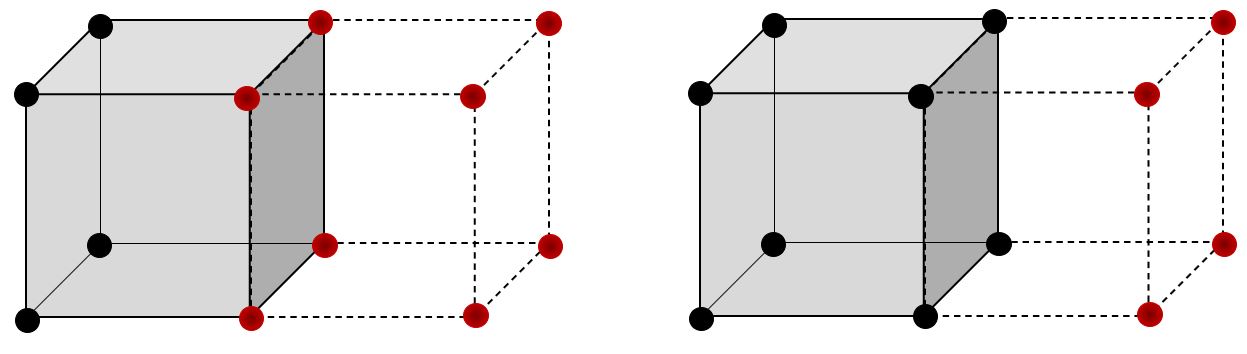}
\caption{Left: full Dirichlet loading, right: partial Dirichlet loading. Red nodes are loaded upon activation, nodes with pre-existing nodal solution indicated in black}
\label{fig:Dirichlet}
\end{figure}

\begin{figure}
\centering
\includegraphics[scale=0.33]{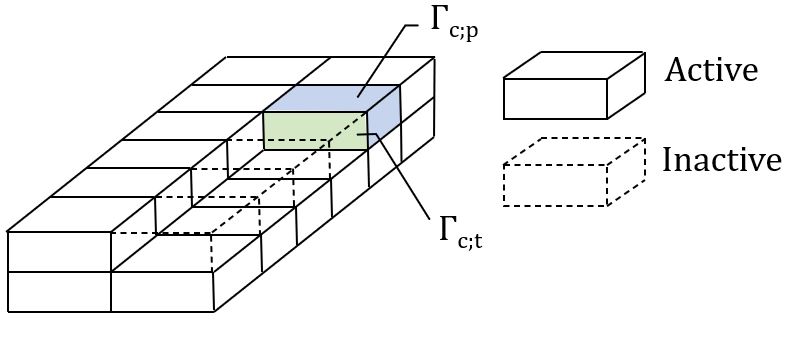}
\caption{Application of convective boundary conditions at element surfaces}
\label{fig:bc_el}
\end{figure}

\indent Upon element activation, the material deposition was simulated by prescribing the extrusion temperature directly at the nodes of the activated elements; i.e. the load was applied as a Dirichlet boundary condition. The load was prescribed at all nodes of the activated element as opposed to only on those nodes without a solved degree of freedom from the previous time step (figure \ref{fig:Dirichlet}). Even though prescribing the temperature at all nodes could result in convergence issues due to the overwriting of the existing solution at certain nodes, this wasn't encountered in any of the simulations presented in this work (section \ref{sec:results}). Moreover, for 8-node thermal finite elements (as used in this work) partial loading would result in only one loaded node in case of adjacent elements which could lead to underestimation of the introduced thermal energy. Thus, this option was not applied. \\
\indent In most FFF printing applications the effects of radiation are negligible as the effect of convection is governing \cite{Costa2015}. Thus, the Neumann boundary condition representing convective heat transfer can be expressed as follows:

\begin{equation}
K_0\nabla T(\textbf{x},t)\mathbf{\cdot n}+h(T_b-T_{\infty})=0 \quad \mathbf{x} \in \Gamma_c\\
\label{eq:bc_Neumann}
\end{equation}

\noindent where $\Gamma_c$ are the free surfaces of the activated elements. Distiction is made between the free surfaces located at the edges of the geometry $\Gamma_{c;p}$ and the temporary free surfaces which are surfaces adjacent to inactive elements $\Gamma_{c;t}$ (figure \ref{fig:bc_el}). During the transient analysis, the latter were continuously updated and identified. The heated printing bed was modeled by prescribing a fixed temperature at the bottom face of the geometry $\Gamma_b$:

\begin{equation}
T(\textbf{x},t)=T_b \quad \textbf{x}\in \Gamma_b
\label{eq:bc_Dirichlet}
\end{equation}

\subsection{Efficiency Measures}
\subsubsection{Remeshing}
\begin{figure*}
\centering
\subfloat[Final discretized geometry]{\includegraphics[scale=0.5]{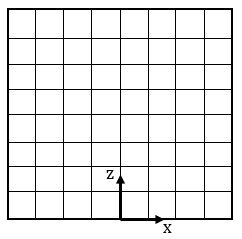}}\hspace{1em}
\subfloat[Quiet activation method; active elements in straight lines, inactive elements in dashed lines]{\includegraphics[scale=0.5]{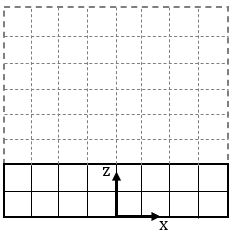}}\\
\subfloat[Hybrid activation method; subsequent remeshing steps]{\includegraphics[scale=0.5]{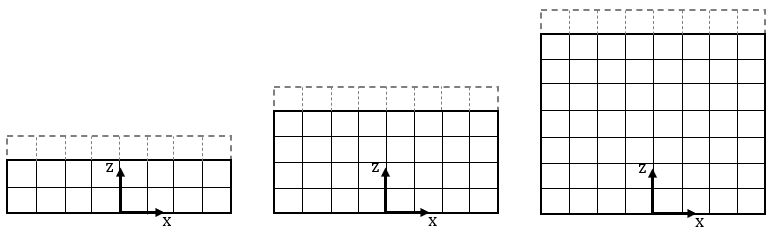}}
\caption{Various element activation methods }
\label{fig:elem_act}
\end{figure*}

In order to avoid having all dofs present during all time steps required to solve the full geometry, Denlinger et al. proposed a hybrid element activation method \cite{Denlinger2014}.Instead of solving the full mesh with many inactive layers in one simulation, often referred to as the quiet activation method, remeshing occurs a predefined number of times until the final geometry is fully solved. Both activation methods are shown schematically in figure \ref{fig:elem_act}. Within each remeshing step in the hybrid activation approach, a user-defined number of inactive or quiet layers nh$_{\text{add}}$ are added to the mesh and then sequentially activated. The number of times remeshing must occur will depend on nh$_{\text{add}}$.\\
\indent In order to ensure continuity of the solution between the meshes of two subsequent remeshing steps, the solution of the previous mesh must be mapped onto the newly discretized geometry prior to the continuation of the analysis. The nodes within the coinciding part of the geometry of two subsequent meshes will be assigned the solution or nodal temperature of the previous remeshing step as an initial condition. The nodes in the quiet layers will be assigned the ambient temperature $T_a$. Therefore, the initial condition is expressed as:

\begin{equation}
T(\textbf{x},0)=T_a \quad \textbf{x}\in \Omega_q
\label{eq:IC_quiet}
\end{equation}

\noindent where $\Omega_q$ represents the quiet part of the discretized domain. In the active part of the domain $\Omega_a$, the initial condition can be expressed as:

\begin{equation}
T^{i}(\mathbf{x},0)=T^{i-1}(\mathbf{x},t_e) \quad \mathbf{x} \in \Omega_a
\label{eq:IC_active}
\end{equation}

\noindent The temperatures at $t=0$ in the current remeshing step $i$ are equal to those at the last time step $t_e$ of the previous remeshing step $i-1$. 

\subsubsection{Adaptive Coarsening}
\label{subsec:adapt_coars}
\begin{figure}
\centering
\subfloat[CF=2]{\includegraphics[scale=0.35]{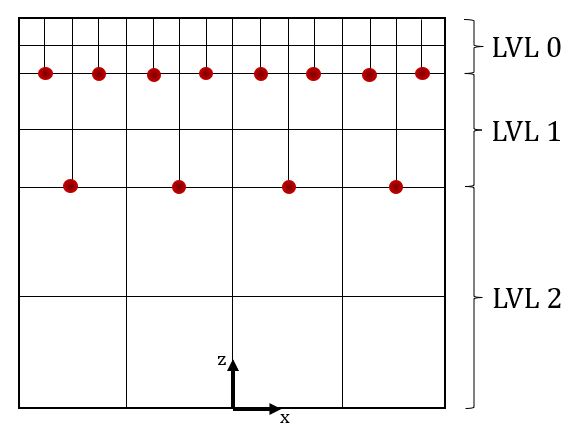}}\hspace{1em}
\subfloat[CF=3]{\includegraphics[scale=0.35]{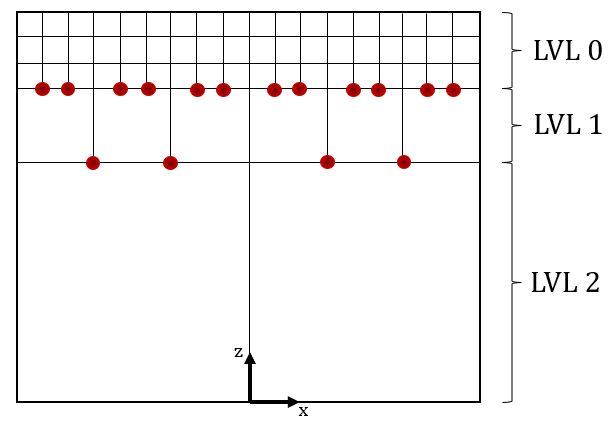}}
\caption{Similarly sized geometries with gradual coarsening patterns (varying CF). Hanging nodes shown in red.}
\label{fig:grad_coars}
\end{figure}

The coarsening strategy in this work consists of a gradual coarsening approach: the elements become increasingly larger as the distance to the printing nozzle increases. How the coarsening is done is determined by the coarsening pattern and the moment at which coarsening occurs is determined by the coarsening condition as the coarsening is done adaptively.\\
\indent The coarsening pattern is controlled by two parameters, namely the number of coarsening levels MLVL and the coarsening factor CF which determines how many elements are merged from one coarsening level to another. Figure \ref{fig:grad_coars} shows an example of a 2D coarsened geometry with two levels of coarsening and varying values of CF. CF=2 was applied in all coarsened meshes in this work and thus two elements in x-, two elements in y- and two elements in z-direction were merged to form one coarse element in the subsequent coarsening level. Coarsening will inevitably lead to hanging nodes (figure \ref{fig:grad_coars}) at the interface between fine and coarse layers and/or between coarse layers of different coarsening levels. These nodes are attached to the elements above the interface layer, but not to the mesh below the interface layer. A continuous solution along this coarser element during the analysis is ensured by defining kinematic constraints prior to solving the global system of equations.

\begin{figure}
\centering
\subfloat[Mesh containing 18 elements in x-direction in finest layer.\label{fig:18x18}]{\includegraphics[scale=0.33]{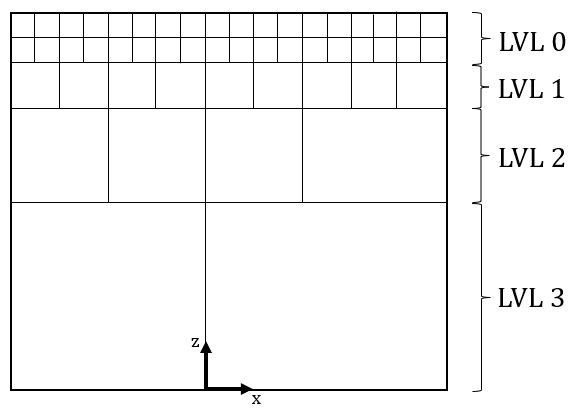}}\hspace{1em}
\subfloat[Mesh containing 19 elements in x-direction in finest layer.\label{fig:19x19}]{\includegraphics[scale=0.33]{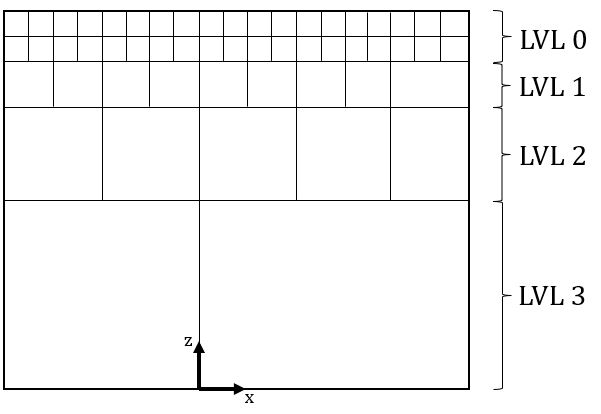}}
\caption{Coarsening patterns in meshes with uneven element configurations}
\label{fig:uneven}
\end{figure}

Figure \ref{fig:grad_coars} shows even meshes where the elements in all coarsening levels are dividable by $\text{CF}^{\text{MLVL}}$. This is usually not the case for most meshes and it is accounted for by slightly changing the coarsening pattern in x- and y-direction such that the number of hanging nodes are minimized. The first level in which an uneven number of elements needs to be coarsened will contain one coarse element that merges an uneven number of elements. The maximum size of this element (length or width) is:
\begin{equation}
l_{\text{max}} \leq 1.5\cdot2^k\cdot l
\end{equation}
where $l_{\text{max}}$ is the maximum length of an element and where $k$ is coarsening level in which the element is coarsened. All neighbouring elements follow CF=2. This can be seen in LVL 2 of figure \ref{fig:18x18} and LVL 1 of figure \ref{fig:19x19}. If the condition isn't satisfied, the uneven element from the previous level is transferred to the next coarsening level without any merging with other elements (LVL 2 in figure \ref{fig:19x19}). This also means that there won't be any hanging nodes at the location of the deviating element in the interface layer.

\begin{figure}
\centering
\subfloat[Current mesh remeshing step $i$]{\includegraphics[width=0.4\linewidth]{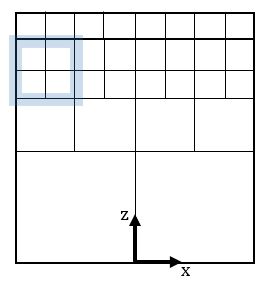}}\hspace{1em}
\subfloat[Potentially coarsened mesh]{\includegraphics[width=0.385\linewidth]{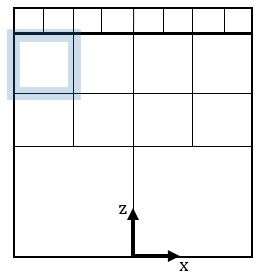}}\\
\subfloat[Domain coarsened element in current mesh: corner nodes in black, interpolated nodes in white \label{fig:domain_element}]{\includegraphics[scale=0.45]{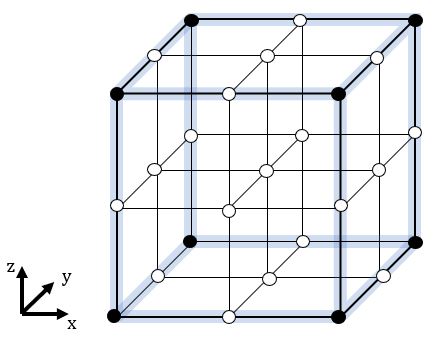}}
\caption{Adaptive coarsening: comparison between current mesh and potential mesh}
\label{fig:coars_check}
\end{figure}

\indent Coarsening occurs when the temperature difference at all nodes between the meshes of two subsequent remeshing steps is smaller than a user-defined threshold. The coarsening condition is checked for each node in the layer that is up for coarsening in the current mesh in remeshing step $i$ and its potentially coarsened mesh. Comparison with the potential mesh will determine if coarsening is appropriate before actually doing so in remeshing step $i+1$. The element configuration in the potential mesh is identical to that of the current mesh, except for the next layer that is up for coarsening (figure \ref{fig:coars_check}). Coarse element dimensions are prematurely assigned to the elements in this layer. The coarsening condition can be expressed as follows:

\begin{equation}
\bigg\lvert \frac{T_j-\hat{T}_j}{T_j} \bigg\rvert<\varepsilon
\label{eq:epsilon}
\end{equation} 

\noindent where $T_j$ is the temperature at node $j$ in the current mesh and where $\varepsilon$ is the user-defined coarsening threshold. $\hat{T_j}$ is the temperature at the projected location of node $j$ in the potential mesh. As this node is not present in the potential mesh, $\hat{T_j}$ must be calculated by linearly interpolating the nodal temperatures of the eight corner nodes of the coarse element (figure \ref{fig:domain_element}). $\varepsilon$ can be defined to be as small as the user deems fit. The influence of $\varepsilon$ will be investigated in section \ref{sec:results}. \\
\indent If all nodes in the potentially coarsened layer satisfy the coarsening condition, coarsening of that layer is appropriate. The same process can be repeated for the other layers in the various coarsening levels. As soon as the coarsening condition isn't met, the nodal coordinates and temperatures from the last approved potential mesh in remeshing step $i$ are saved for mapping in remeshing step $i+1$. The thermal transient analysis in remeshing step $i$ can then be concluded.

\subsubsection{Homogenizing Infill Geometries}
\label{subsec:infill}
\begin{figure}
\centering
\includegraphics[scale=0.5]{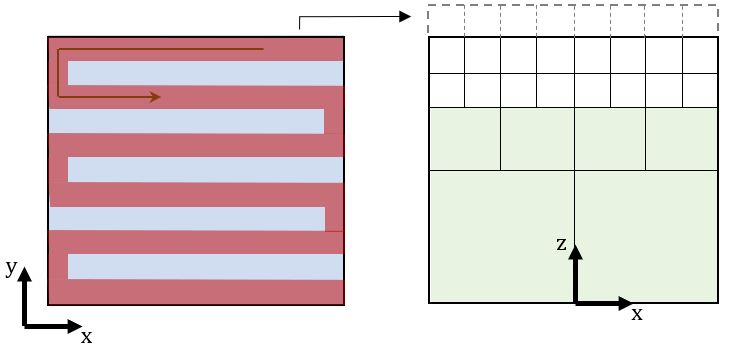}
\caption{left: Layer with an infill pattern (polymer in red and air in blue), right: effective material properties in coarsened layers (indicated in green)}
\label{fig:infill}
\end{figure}

An example of a layer with a rectilinear infill pattern with air content is schematically shown in figure \ref{fig:infill}. When simulating the deposition of such a layer air elements must be activated in addition to the PLA elements. Air elements adjacent to polymer elements are activated simultaneously within the same time step, since they don't contribute to the actual printing time. Coarsening of elements that meet the coarsening condition can be done in the same manner as described in section \ref{subsec:adapt_coars}. The main difference is that elements that are to be merged can consist of only air, only polymer or a combination of both materials. It was found in \cite{Ramos2022} that when modeling geometries with complex, air-filled infill structures, accurate heat transfer can be simulated with simplified infill structures as long as the infill density is respected. Thus, instead of considering the exact material configuration of the merged elements, effective or homogenized material properties are assigned to coarsened elements in which the influence of the infill density $\alpha$ of the printed part is included (figure \ref{fig:infill}). The exact infill pattern is always respected in the fine layers by exactly following the deposition path of the printing nozzle during the element activation.\\
\indent There are various methods to calculate effective properties of heterogeneous materials and porous media \cite{Pietrak2015}. In this work a simple approach is chosen to calculate effective values of the relevant thermal properties, i.e. the conductivity and the (volumetric) heat capacity:

\begin{equation}
K_{0;eff}=(1-\alpha)\cdot K_{0;air}+\alpha\cdot K_{0;pol}
\label{eq:effect_cond}
\end{equation}

\begin{equation}
C_{eff}=(1-\alpha)\cdot C_{air}+\alpha\cdot C_{pol}
\label{eq:effect_heatcap}
\end{equation}

\begin{equation}
C_{air}=\rho_{air}\cdot c_{p;air}
\end{equation}

\begin{equation}
C_{pol}=\rho_{pol}\cdot c_{p;pol}
\end{equation}

\begin{equation}
\alpha=\frac{V_{pol}}{V_{air}+V_{pol}}
\label{eq:alpha}
\end{equation}

\noindent where subscripts $_{air}$ and $_{pol}$ refer to air and polymer respectively, where $C$ is the volumetric heat capacity [J/m$^3$ K] and where $V$ is the volume of a material in the printed part. 

\subsection{Numerical set-up}
\label{subsec:mat_proc}

All simulations presented in this work were carried out in Ansys Mechanical APDL V19.2. All geometries were discretized with SOLID70, 3D 8-node thermal finite elements. The thermal properties that were assigned to the elements are listed in table \ref{tab:mat_prop}. These properties were also used to calculate the homogenized material properties assigned to the coarsened elements in the infill geometries. The dimensions of the non-coarsened elements were determined by the geometry of the filaments of the printed geometries. The element width $wf$ was chosen to be equal to the filament width, the element length $dl$ was assumed to be equal to the element width and the element height $dh$ was equal to the layer height.

\section{Results \& Discussion}
\label{sec:results}

\subsection{Experimental results}
\begin{figure}[h]
\centering
\subfloat[Time-temperature plot N1]{\includegraphics[scale=0.45]{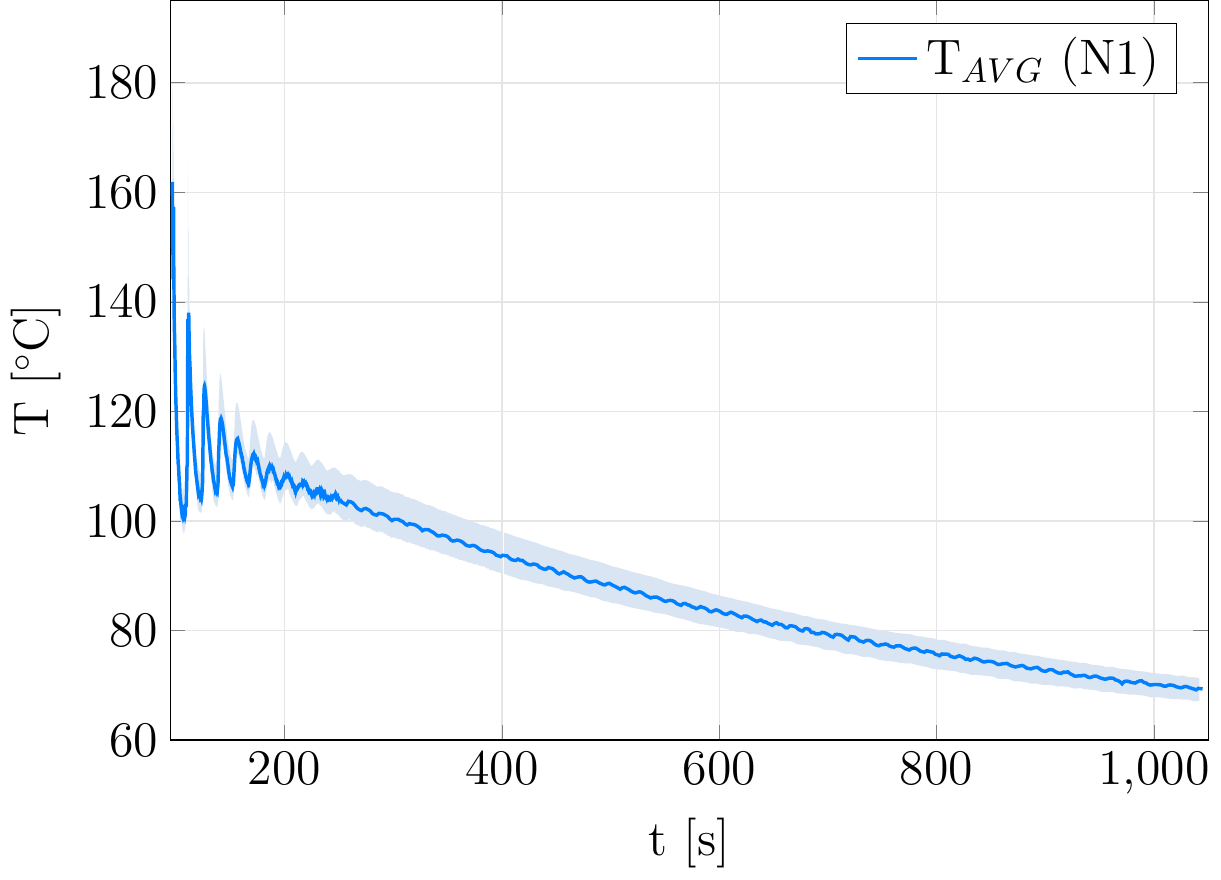}}\hspace{1em}
\subfloat[Time-temperature plot N2]{\includegraphics[scale=0.45]{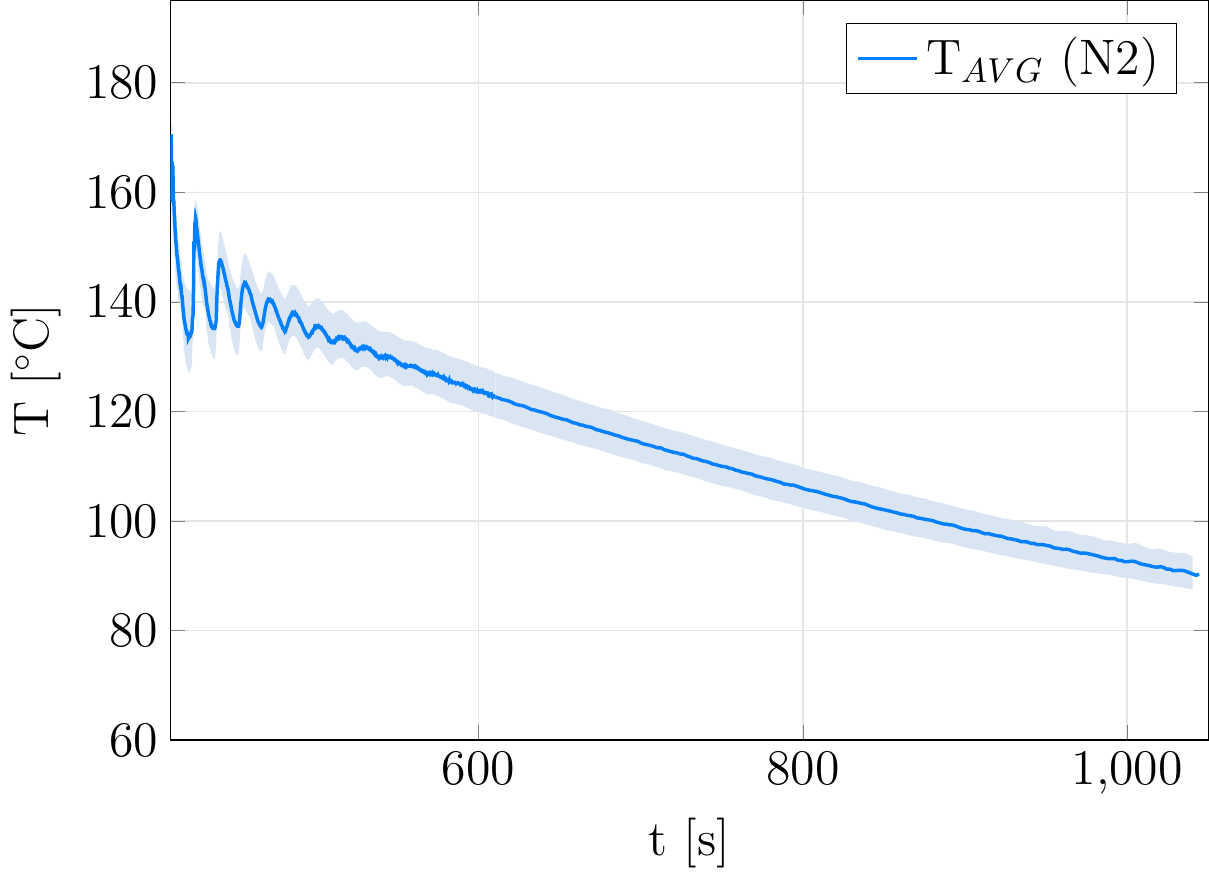}}\\
\subfloat[Time-temperature plot N3]{\includegraphics[scale=0.45]{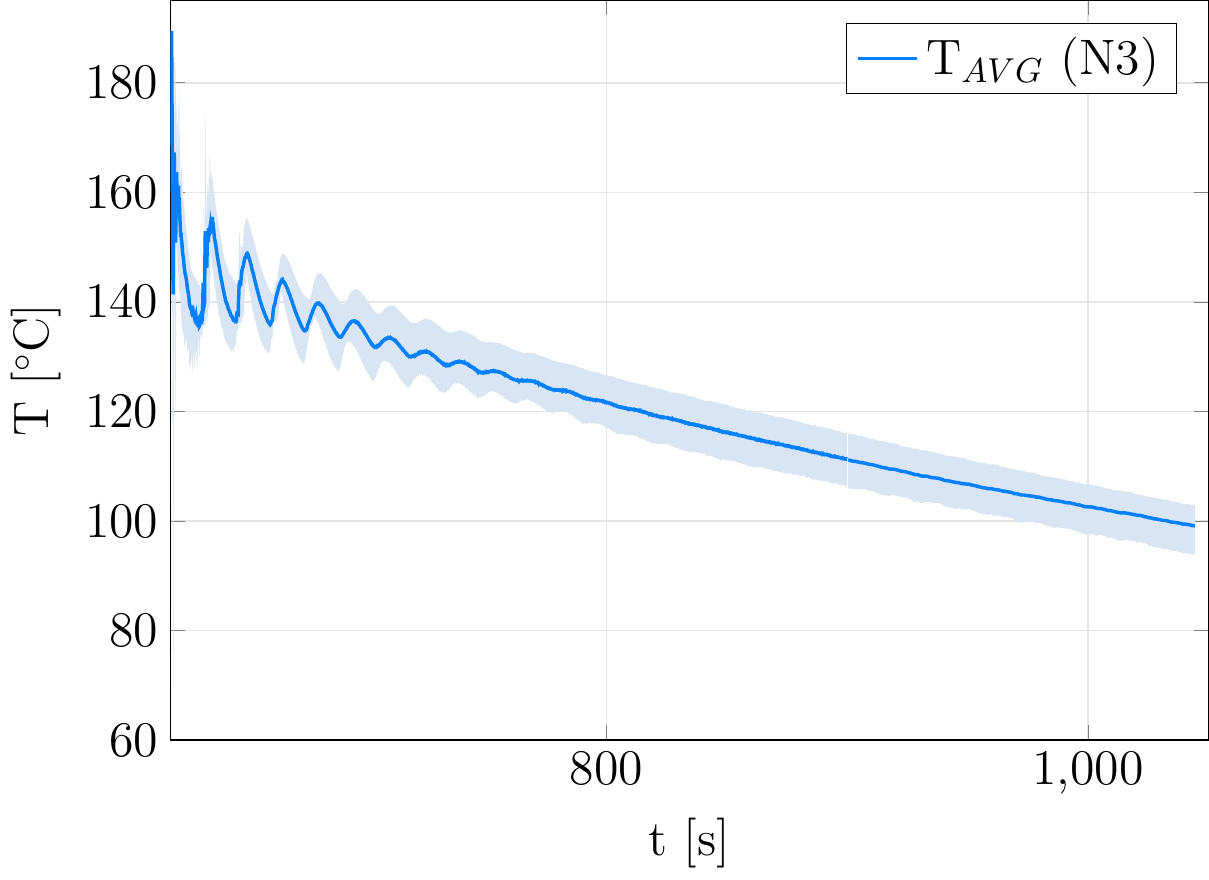}}
\caption{Experimental thermal measurements at various measuring locations}
\label{fig:exp_res}
\end{figure}

Figure \ref{fig:exp_res} shows the results of the thermal measurements performed with the thermocouples during the printing of the block at the three measuring locations N1, N2, N3. For each measuring location, the measured average temperature as well as the temperature envelope are plotted as a function of time. The $T(t)$ graphs don't start at $t=0$ which would be the start of the printing process, but at the time at which the filament is deposited on top of the thermocouple. \\
\indent It can be seen that the measured deposition temperature, captured by the first peak in the $T(t)$ graph, does not equal the prescribed nozzle temperature $T_n$ of 210$^{\circ}$C at any of the measuring locations. The temperature of the deposited filament was repeatedly measured to be significantly lower than the nozzle temperature. This observation was also made in \cite{Vanaei2021a}. Thus, instead of using the nozzle temperature in the numerical simulations, the average deposition temperature from all the experimental measurements of 175$^{\circ}$C was used as the activation temperature.

\subsection{Validation simulation framework}
\label{subsec:valid}
Before experimentally validating the simulation framework, the validity of the simulation framework is presented by investigating the effect of the efficiency measures on the numerical accuracy. 

\begin{figure}[H]
\centering
\subfloat[Evolution of the number of degrees of freedom over time]{\includegraphics[scale=0.45]{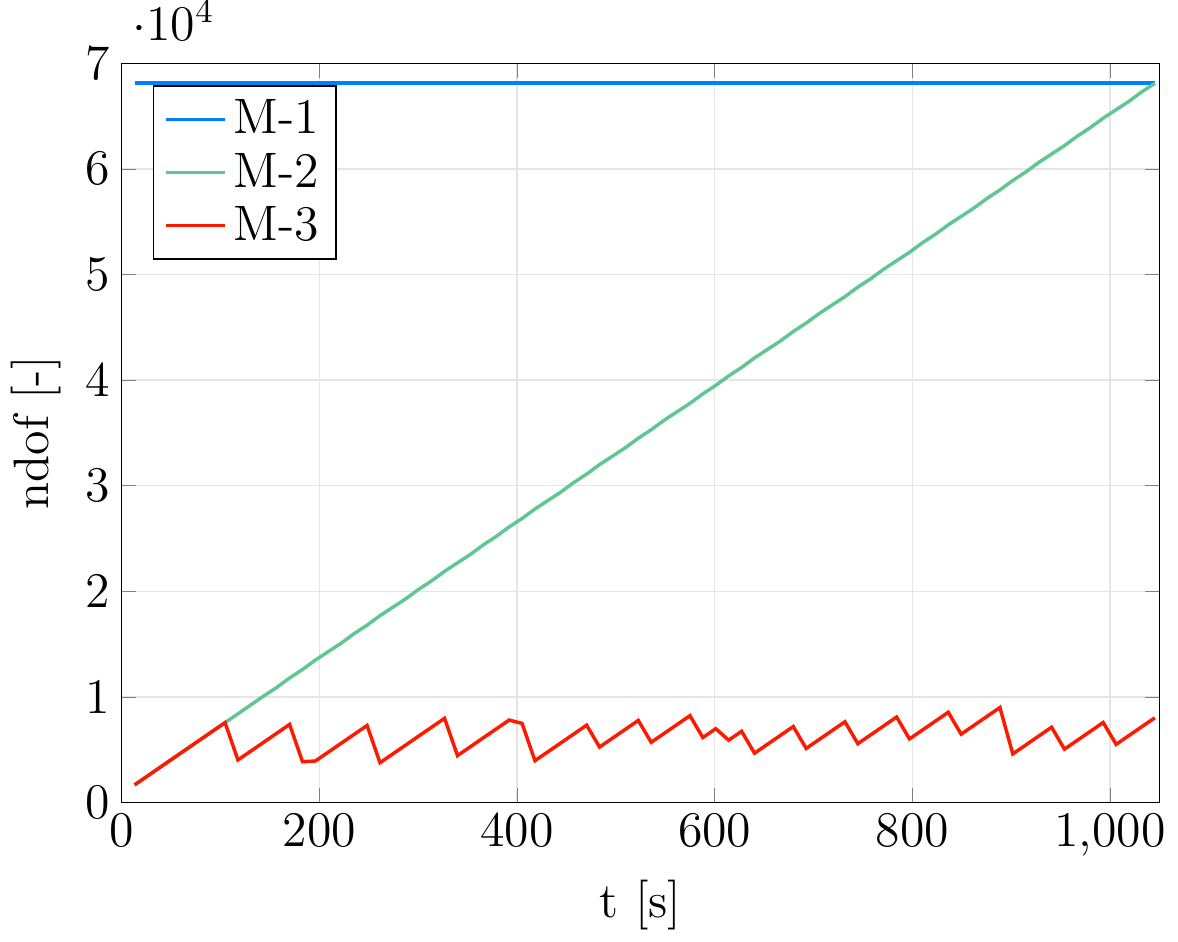}}\hspace{1em}
\subfloat[Time-temperature plot node 1]{\includegraphics[scale=0.45]{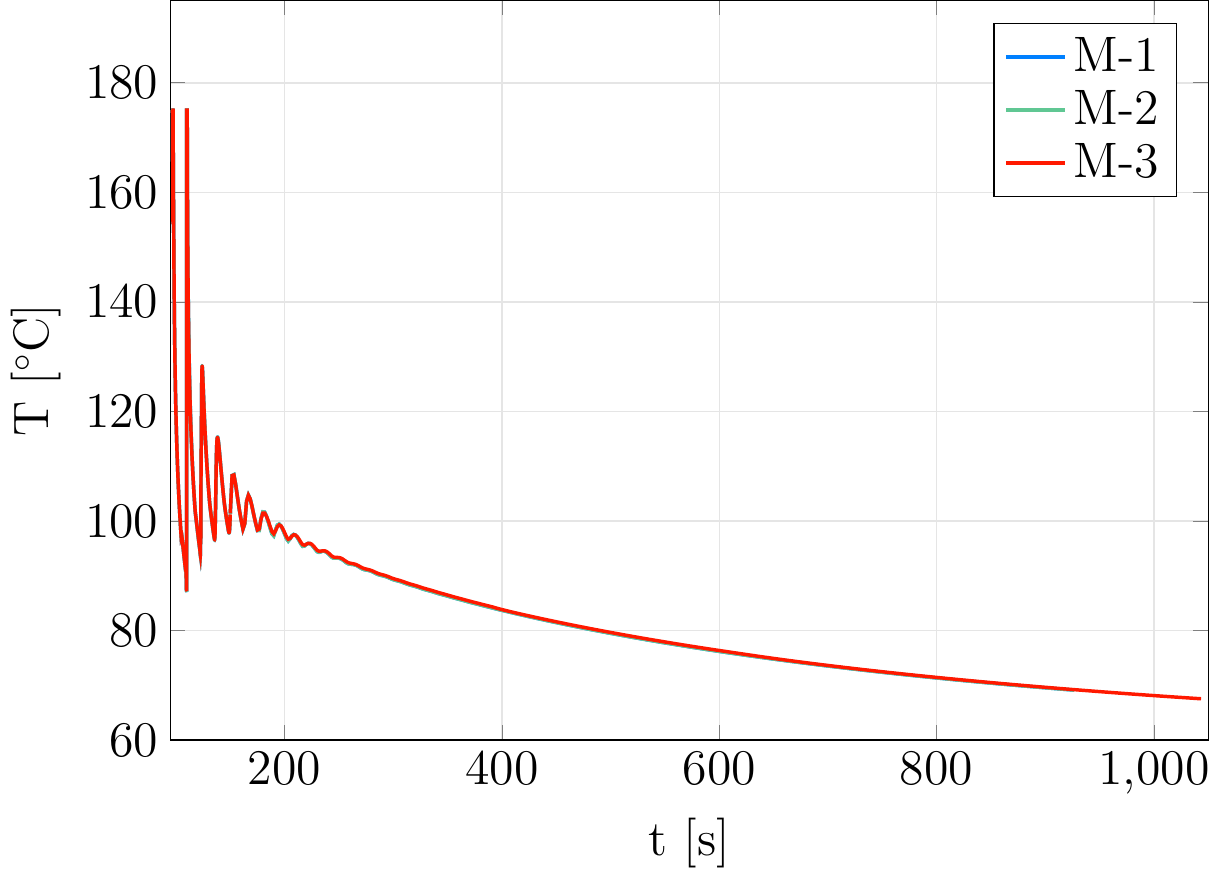}}\\
\subfloat[Time-temperature plot node 2]{\includegraphics[scale=0.45]{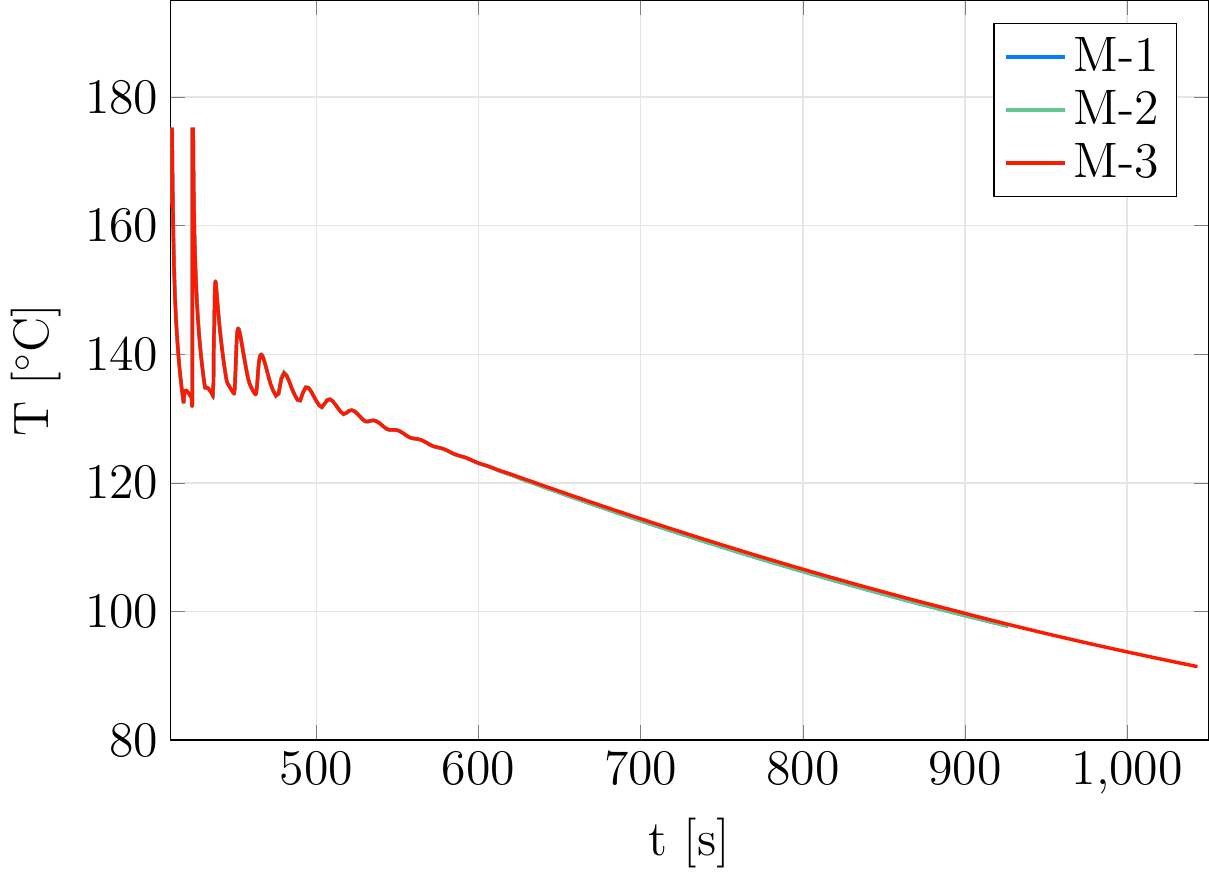}}\hspace{1em}
\subfloat[Time-temperature plot node 3]{\includegraphics[scale=0.45]{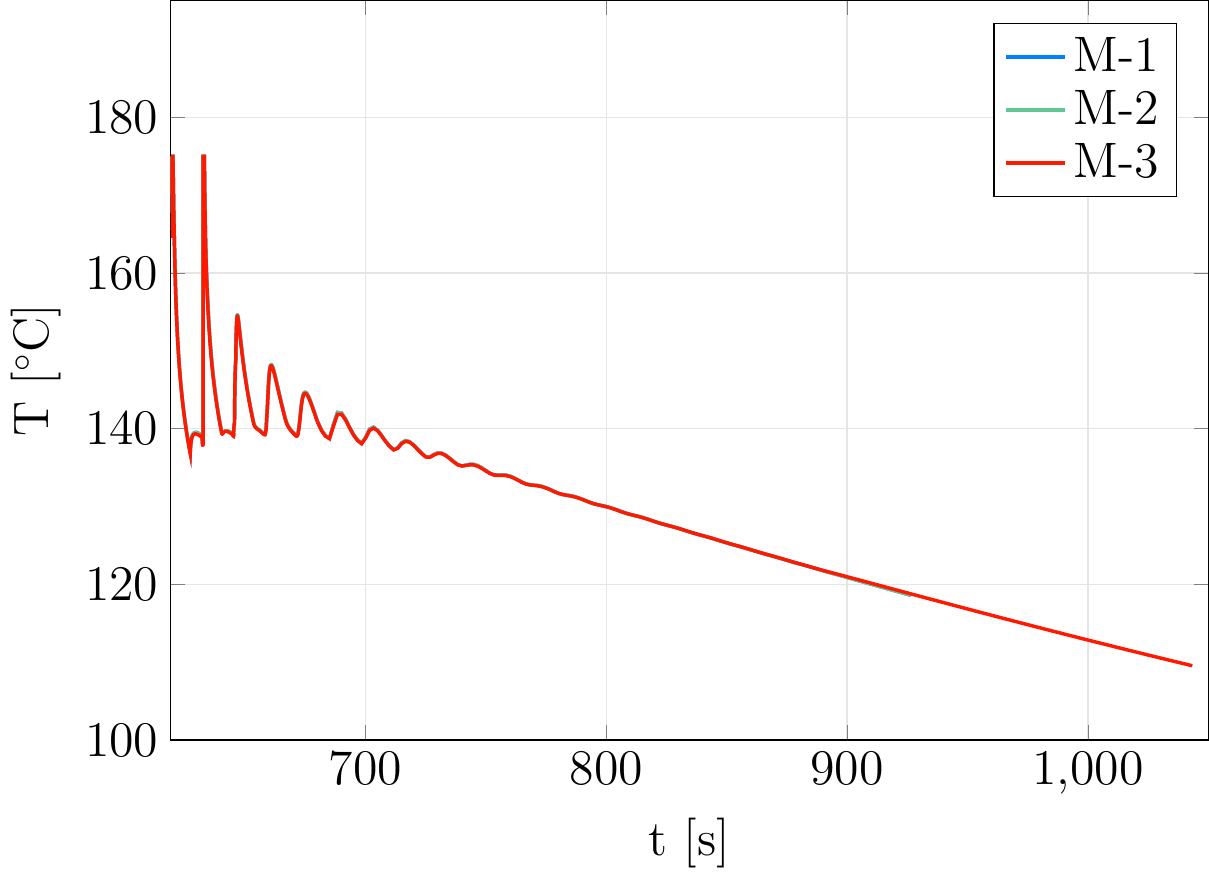}}
\caption{Influence of remeshing and coarsening: M-1 vs. M-2 vs. M-3}
\label{fig:results_def}
\end{figure}

\begin{figure}[H]
\centering
\subfloat{\includegraphics[width=0.31\linewidth]{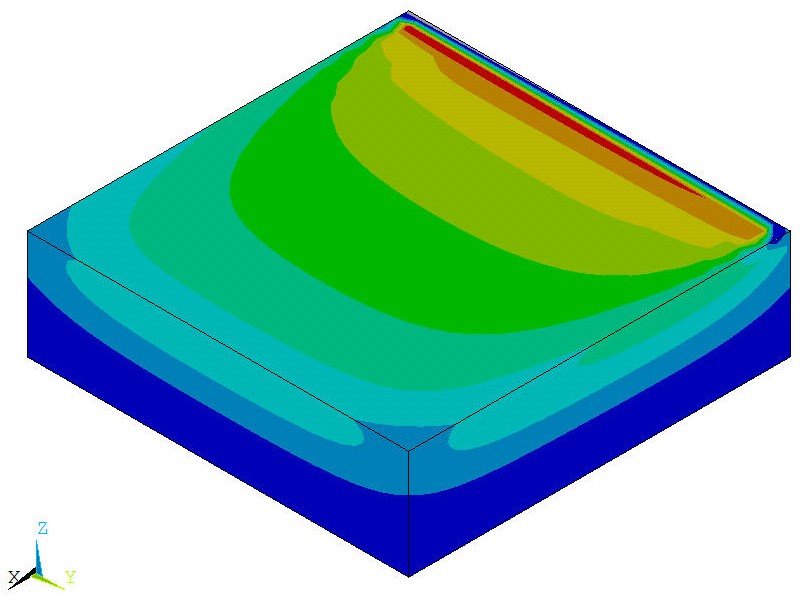}}\hspace{1em}
\subfloat{\includegraphics[width=0.3\linewidth]{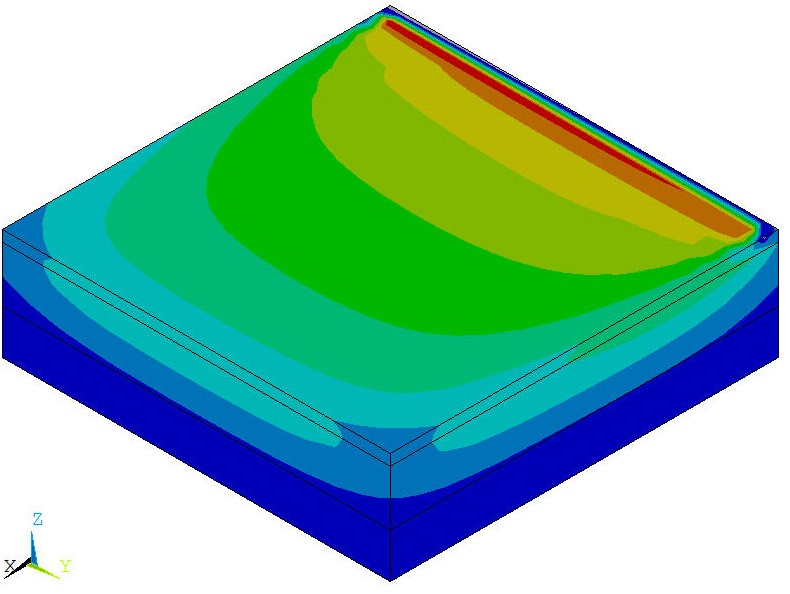}}\\
\subfloat{\includegraphics[width=0.31\linewidth]{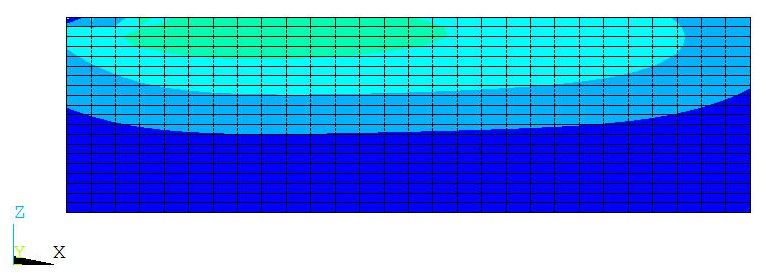}}\hspace{1em}
\subfloat{\includegraphics[width=0.3\linewidth]{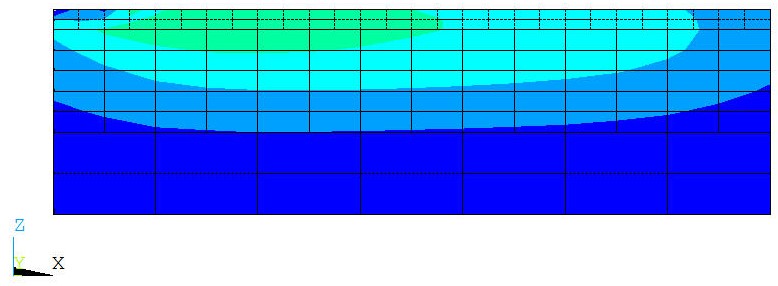}}\\
\subfloat{\includegraphics[scale=0.32]{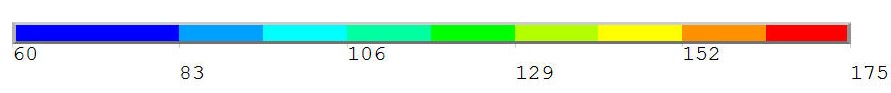}}
\caption{Temperatures after solving 25\% of the geometry. Left: Fine mesh (M-1), right: coarsened mesh (M-3). Cross-sectional plane at  y=$\frac{4}{7}\cdot$l}
\label{fig:def_no19}
\end{figure}

\begin{figure}[H]
\centering
\subfloat{\includegraphics[width=0.31\linewidth]{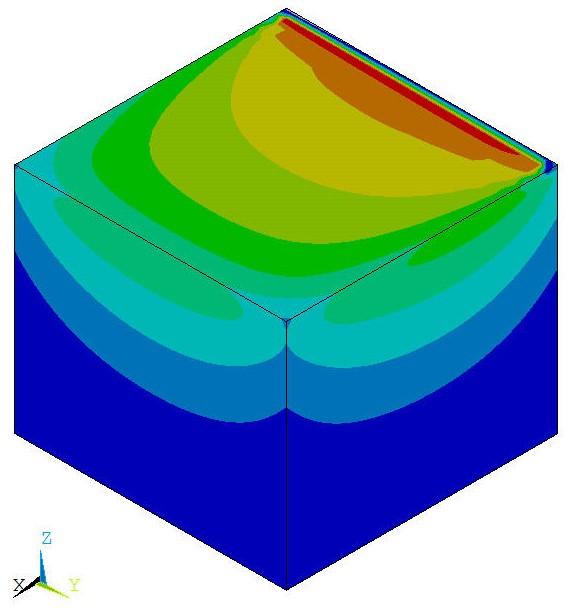}}\hspace{1em}
\subfloat{\includegraphics[width=0.3\linewidth]{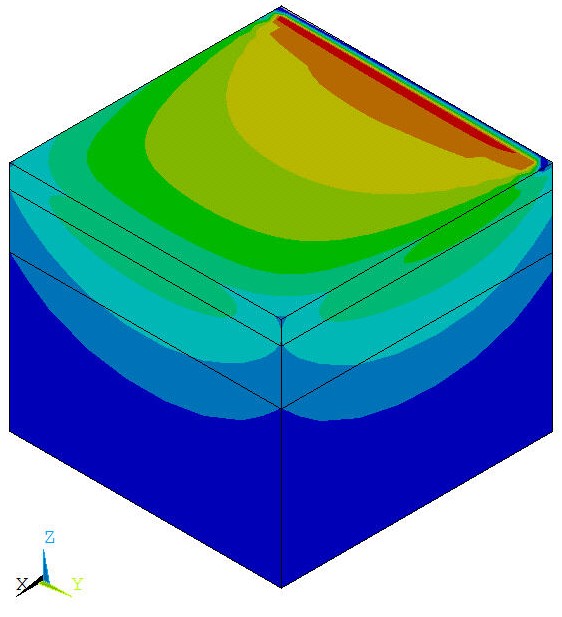}}\\
\subfloat{\includegraphics[width=0.31\linewidth]{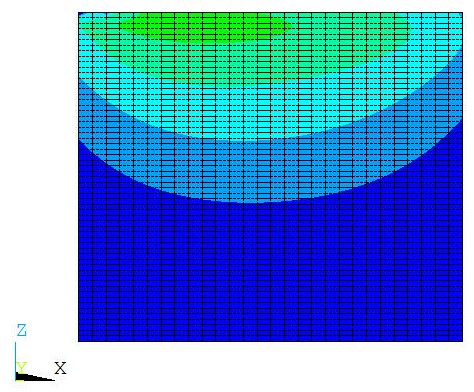}}\hspace{1em}
\subfloat{\includegraphics[width=0.3\linewidth]{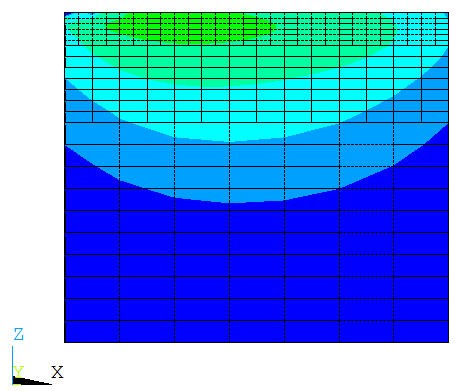}}\\
\subfloat{\includegraphics[scale=0.32]{legend_hor}}
\caption{Temperatures after solving 75\% of the geometry. Left: Fine mesh (M-1), right: coarsened mesh (M-3). Cross-sectional plane at  y=$\frac{4}{7}\cdot$l}
\label{fig:def_no59}
\end{figure}

\paragraph{Numerical validation}
In the first simulation \textbf{M-1}, the block geometry was fully solved with the quiet method, so no remeshing occured between start and finish of the simulation. This is the default simulation to which other results will be compared. In the second simulation the mesh was solved with the hybrid activation method; remeshing occured every time one full layer was activated (\textbf{M-2}). In the third simulation, the mesh was solved with the coarsening framework, thus both remeshing and adaptive coarsening occured (\textbf{M-3}). The default remeshing and coarsening parameters are listed in table \ref{tab:def_param}. The efficiency of the coarsening framework is measured by comparing the total computional time required for the simulations of M-1, M-2 and M-3. The results are listed in table \ref{tab:def_param} and shown in figure \ref{fig:results_def}. \\
\indent A total number of 62,720 time steps was required to solve the full geometry for all of the models. It can be seen the inclusion of remeshing reduced the computational time to 50\% of that of M-1. The application of both remeshing and coarsening reduced the computational time to 19\% of that of the default model. Figure \ref{fig:results_def} shows a linear evolution of the number of dofs over time for M-2 since a fixed number of dofs was added each time the geometry was remeshed. In case of M-1 this number was constant over time as all the dofs were present from the start. When looking at the results for M-3, a clear effect of the adaptive coarsening can be seen; the number of dofs was significantly reduced each time coarsening of the mesh occured. On average the total number of dofs was quite constant and significantly less than in M-1 and M-2.\\
\indent The numerical accuracy of the simulation framework is assessed by comparing the evolution of the temperature over time $T(t)$ for the three different models M-1, M-2 and M-3. The $T(t)$ results are presented in figure \ref{fig:results_def}. There is very good agreement between the three models for all of the measurement points. The model with both remeshing and coarsening is perfectly capable of capturing the (re)heating peaks and cooling of the filament that is displayed by the finest mesh M-1. Thus, at a local level model M-3 agrees well with both M-1 and M-2. A global comparison of the temperature fields has also been made by looking at the temperature contour plots at various moments during the simulation. These are shown in figures \ref{fig:def_no19} \& \ref{fig:def_no59}. For M-1 only the active part of the mesh has been displayed. There is also great agreement between the fine and coarsened mesh in the global temperature profiles. Model M-3 is capable of accurately capturing the correct local and global solution as displayed in the fine mesh without any remeshing and coarsening. 

\begin{table}
\centering
\begin{tabular}{lccccccccc}
\toprule 
& \textbf{M-1} & \textbf{M-2} & \textbf{M-3} & \textbf{M-4} & \textbf{M-5} & \textbf{M-6} & \textbf{M-7} & \textbf{M-8} & \textbf{M-9}\\
\midrule
nh$_{\text{add}}$ & n.a. & 1 & 1 & 2 & 4 & 1 & 1 & 1 & 1\\
CF & n.a. & n.a. & 2 & 2 & 2 & 2 & 2 & n.a. & 2 \\
CLVL & n.a. & n.a. & 3 & 3 & 3 & 3 & 3 & n.a. & 3 \\
$\varepsilon$ & n.a. & n.a. & 0.01 & 0.01 & 0.01 & 0.02 & 0.05 & n.a. & 0.05\\
$\alpha$ & 1 & 1 & 1 & 1 & 1 & 1 & 1 & 0.5 & 0.5\\
\midrule
t$_c$ [-] & 1 & 0.5 & 0.19 & 0.19 & 0.22 & 0.16 & 0.16 & 0.35 & 0.24\\
\bottomrule 
\end{tabular}
\caption{Computational times t$_c$ (relative to default model M-1) for models with varying remeshing \& coarsening parameters}
\label{tab:def_param}
\end{table} 

\begin{figure}[h]
\centering
\subfloat[Validation $T(t)$ N1]{\includegraphics[scale=0.45]{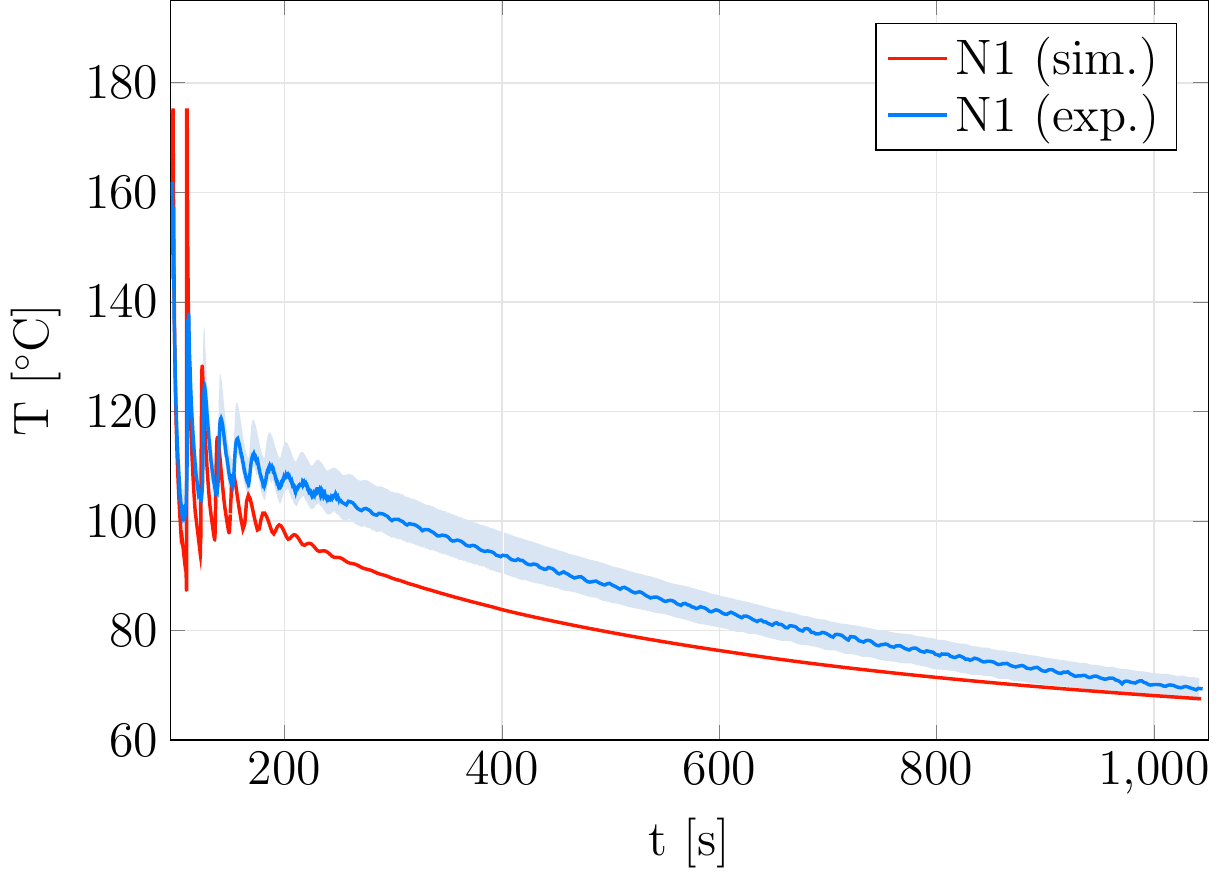}}\hspace{1em}
\subfloat[Validation $T(t)$ N2]{\includegraphics[scale=0.45]{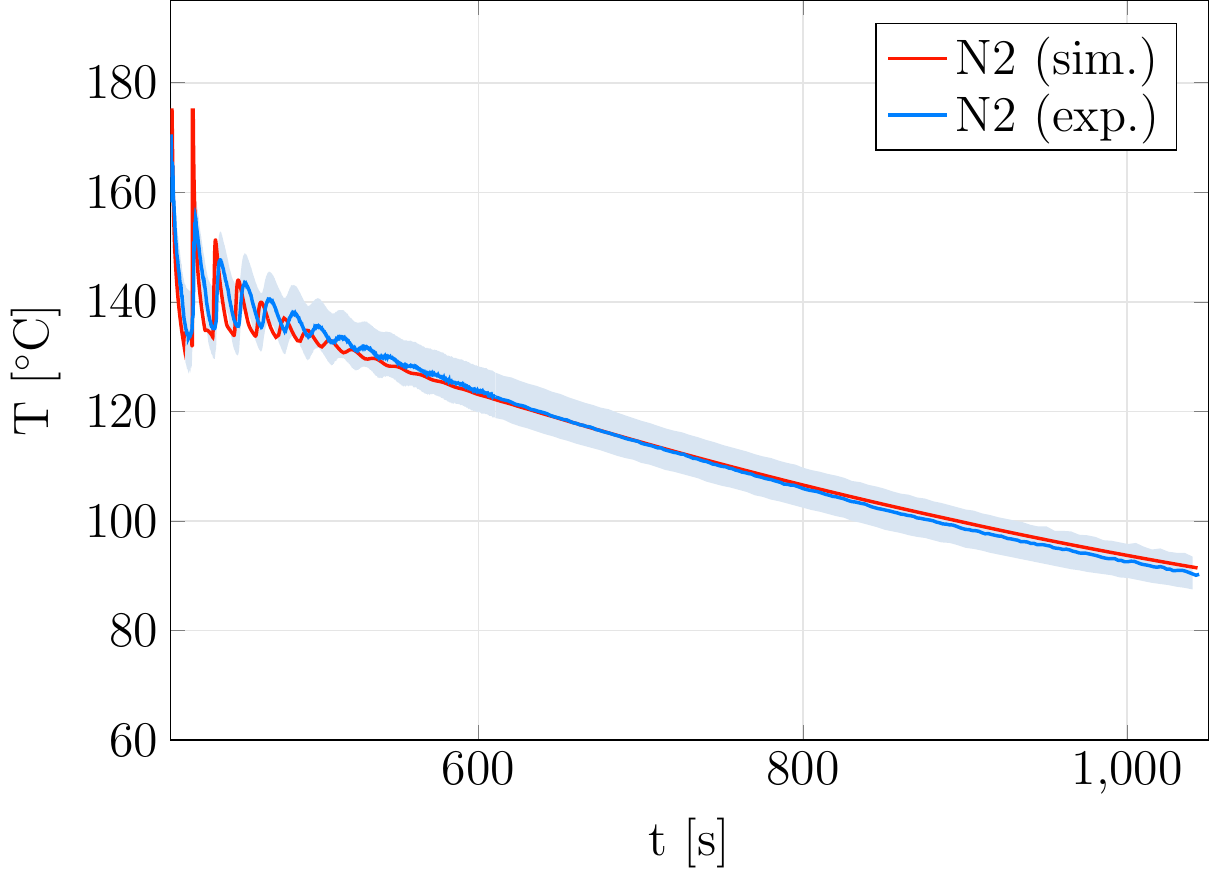}}\\
\subfloat[Validation $T(t)$ N3]{\includegraphics[scale=0.45]{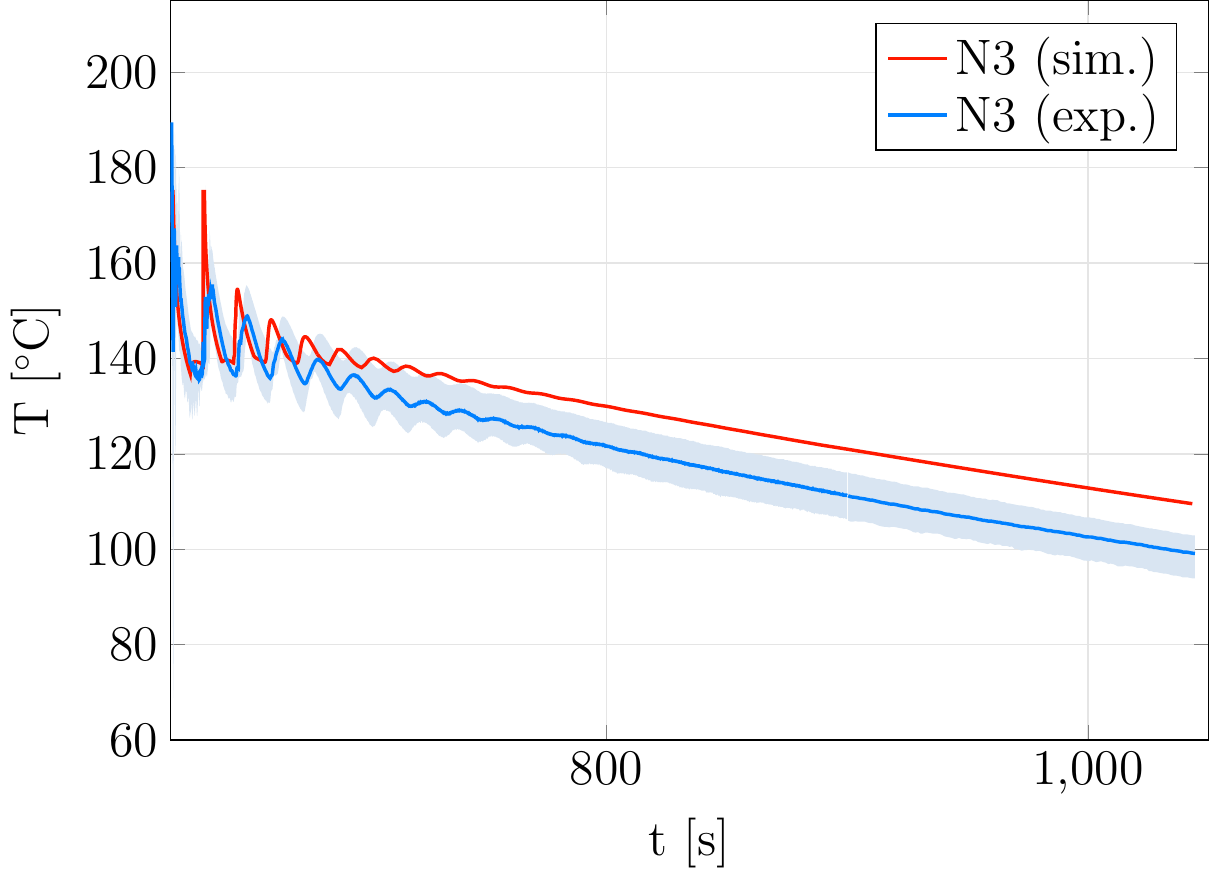}}
\caption{Experimental validation of numerical model \textbf{M-3}}
\label{fig:simexp}
\end{figure}

\paragraph{Experimental validation}
The numerical results generated with model \textbf{M-3} are plotted against the experimental thermal measurements for each measuring point in figure \ref{fig:simexp}. The overall trend in the $T(t)$ evolution is captured well by the simulations. It can be seen that the temperatures are slightly underestimated in the simulation for point N1. This also happens to be the point which is most sensitive to the influence of the thermal boundary conditions as it is located at the bottom part of the printed sample ($z=\frac{1}{10\cdot h}$). There is great agreement between the experimental and numerical results for measuring point N2 and the deviations between the results for N3 are quite small as well ($\approx 5\%$). For all measurement points it can be seen that the temperature at the first and second (re)heating peak is different for the simulation and experiments. The average deposition temperature was used as the element activation temperature in the simulations, thus deviation from the experimental values was to be expected. During the experimental measurements a different deposition temperature was registered at each measurement point. The deviation in the second reheating peak is a direct result of the numerical load application; when an element is activated, the deposition temperature is prescribed at all nodes of that element (section \ref{subsec:mat_dep}). This seems to be an overestimation of what occurs in reality.

\begin{figure}[H]
\centering
\subfloat[Time-temperature plot node 1]{\includegraphics[scale=0.45]{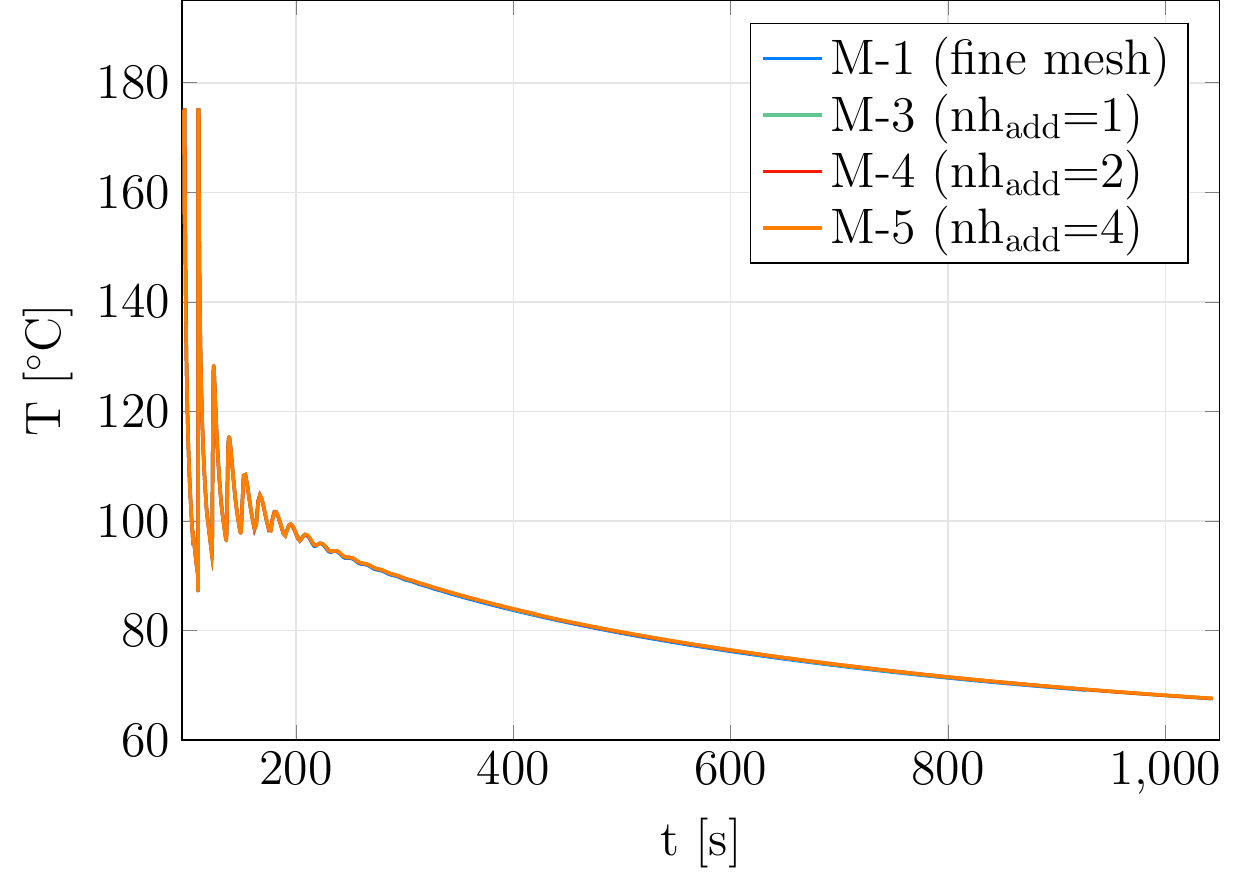}}\hspace{1em}
\subfloat[Time-temperature plot node 2]{\includegraphics[scale=0.45]{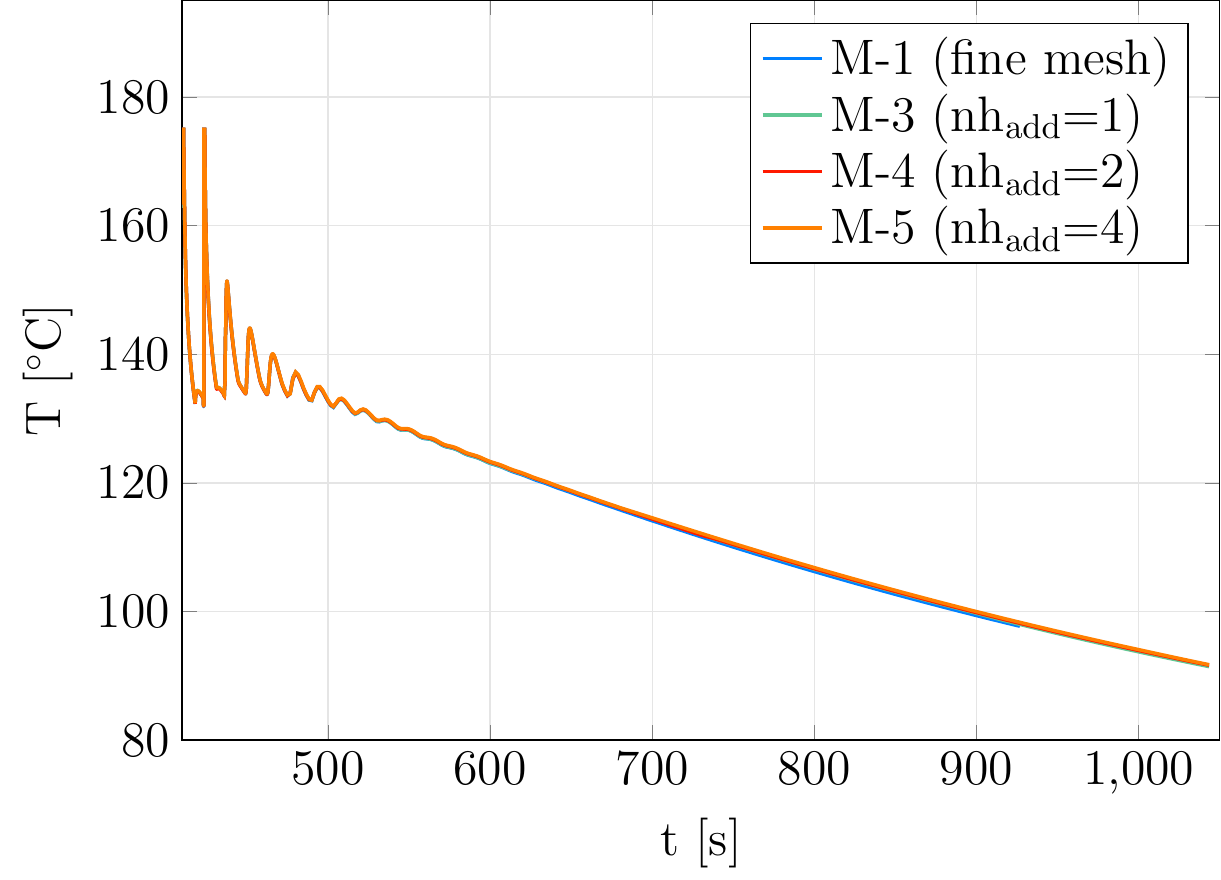}}\\
\subfloat[Time-temperature plot node 3]{\includegraphics[scale=0.45]{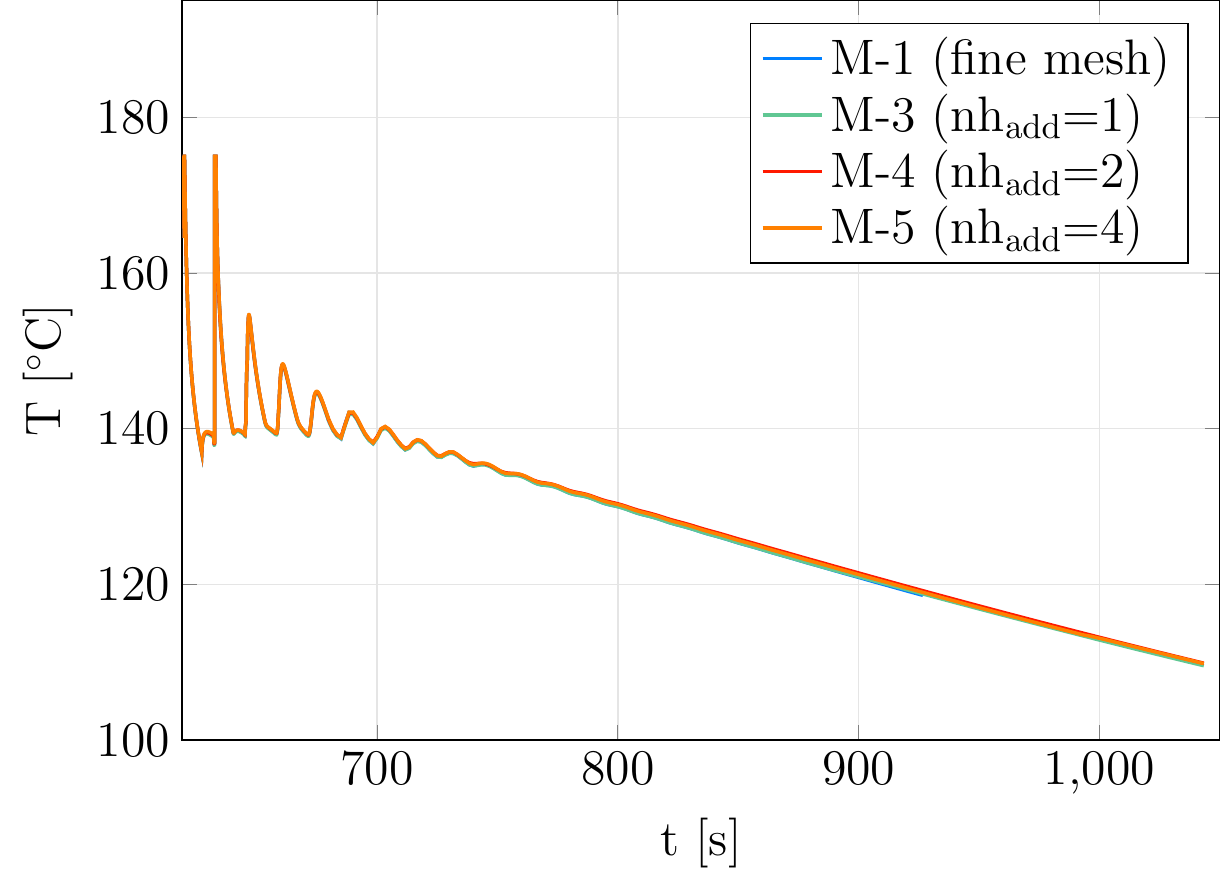}}
\caption{Influence of nh$_{\text{add}}$ }
\label{fig:Tt_nhadd}
\end{figure}

\begin{figure}[H]
\centering
\subfloat[M-3 (nh$_{\text{add}}$=1)]{\includegraphics[width=0.29\linewidth]{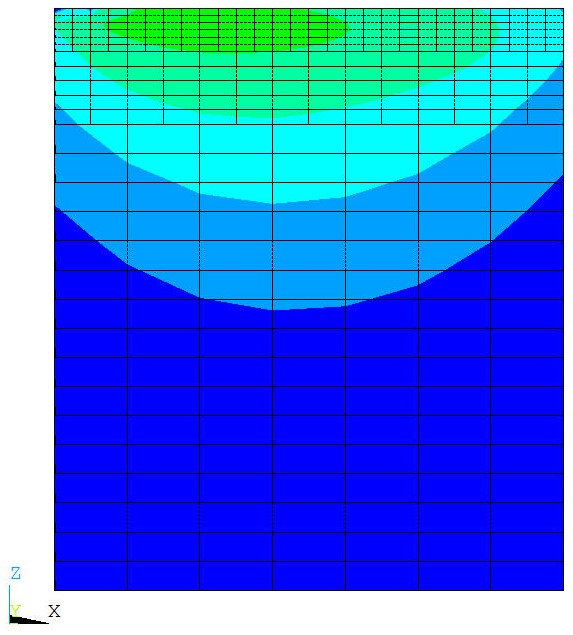}}
\subfloat[M-4 (nh$_{\text{add}}$=2)]{\includegraphics[width=0.285\linewidth]{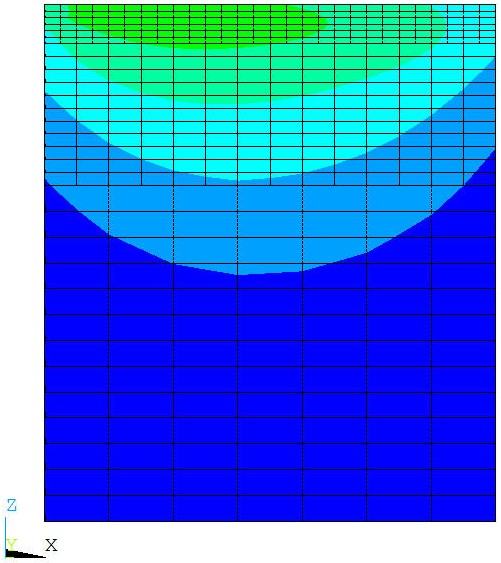}}
\subfloat[M-5 (nh$_{\text{add}}$=4)]{\includegraphics[width=0.28\linewidth]{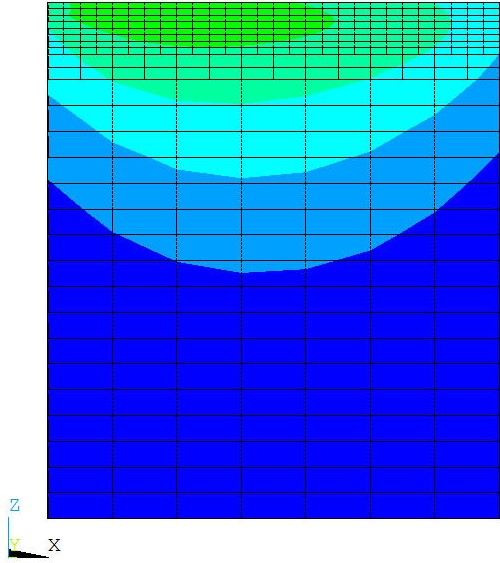}}
\subfloat{\includegraphics[scale=0.3]{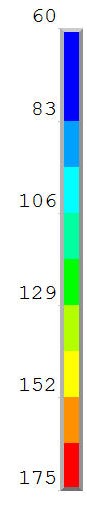}}
\caption{Temperatures in last time step of the final coarsened meshes for models with varying nh$_{\text{add}}$. Cross-sectional plane at  y=$\frac{4}{7}\cdot$l}
\label{fig:contour_nhadd}
\end{figure}

\begin{figure}[H]
\centering
\subfloat[Time-temperature plot node 1]{\includegraphics[scale=0.45]{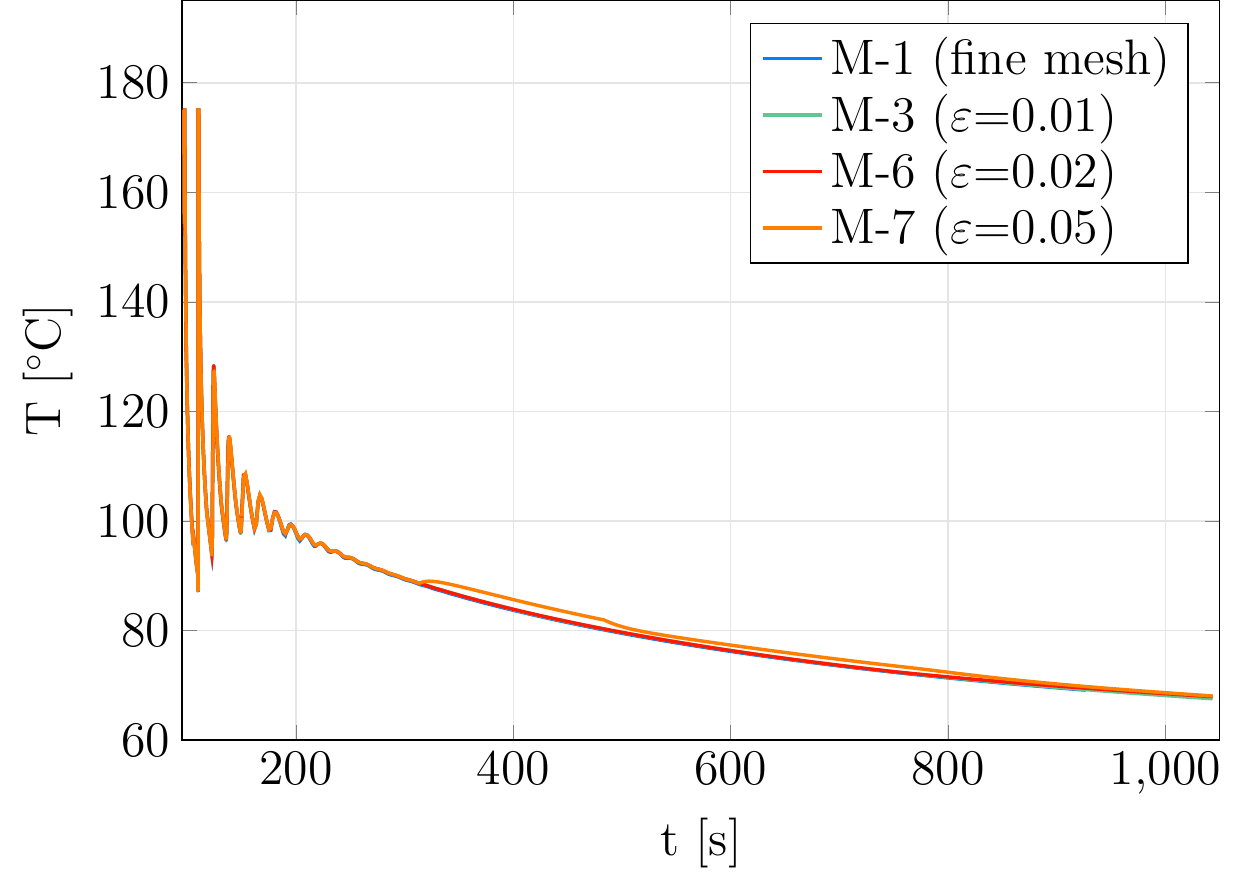}}\hspace{1em}
\subfloat[Time-temperature plot node 2]{\includegraphics[scale=0.45]{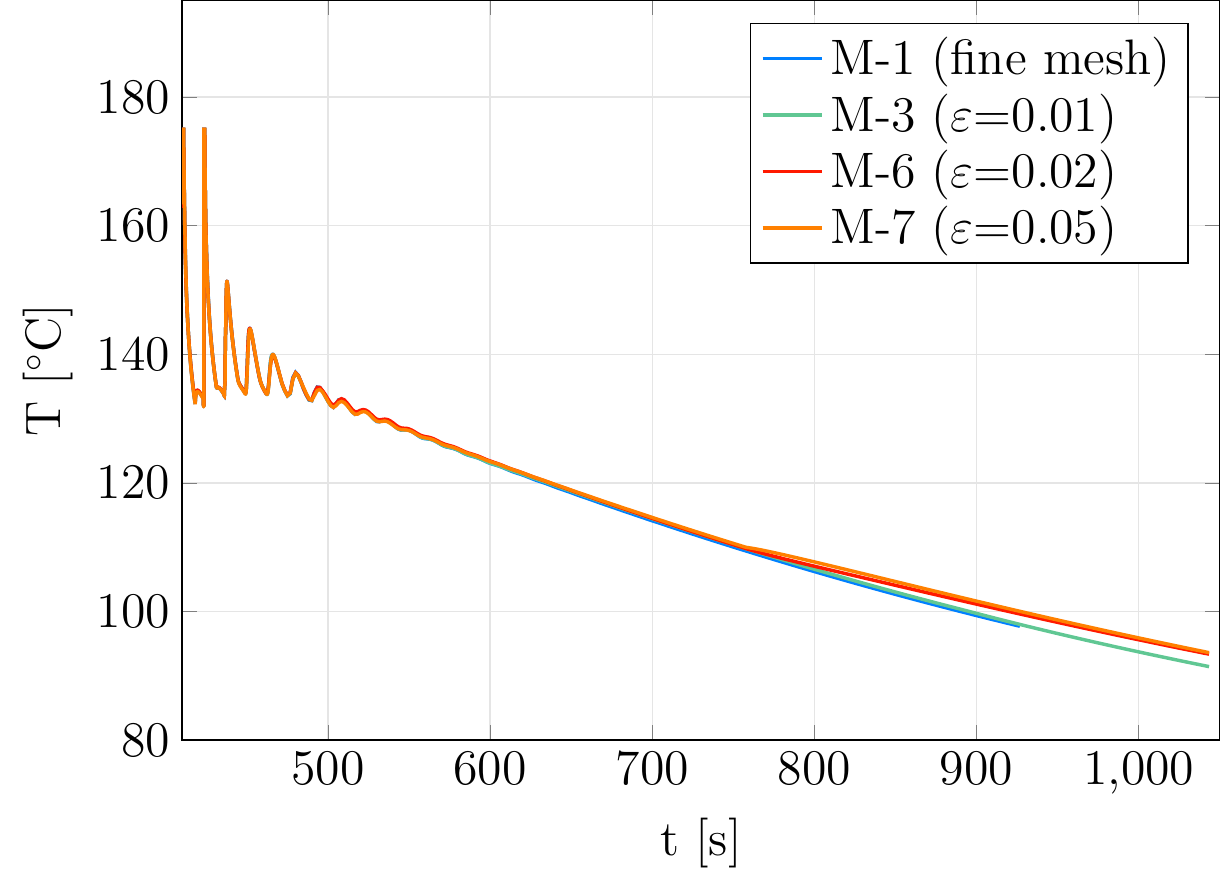}}\\
\subfloat[Time-temperature plot node 3]{\includegraphics[scale=0.45]{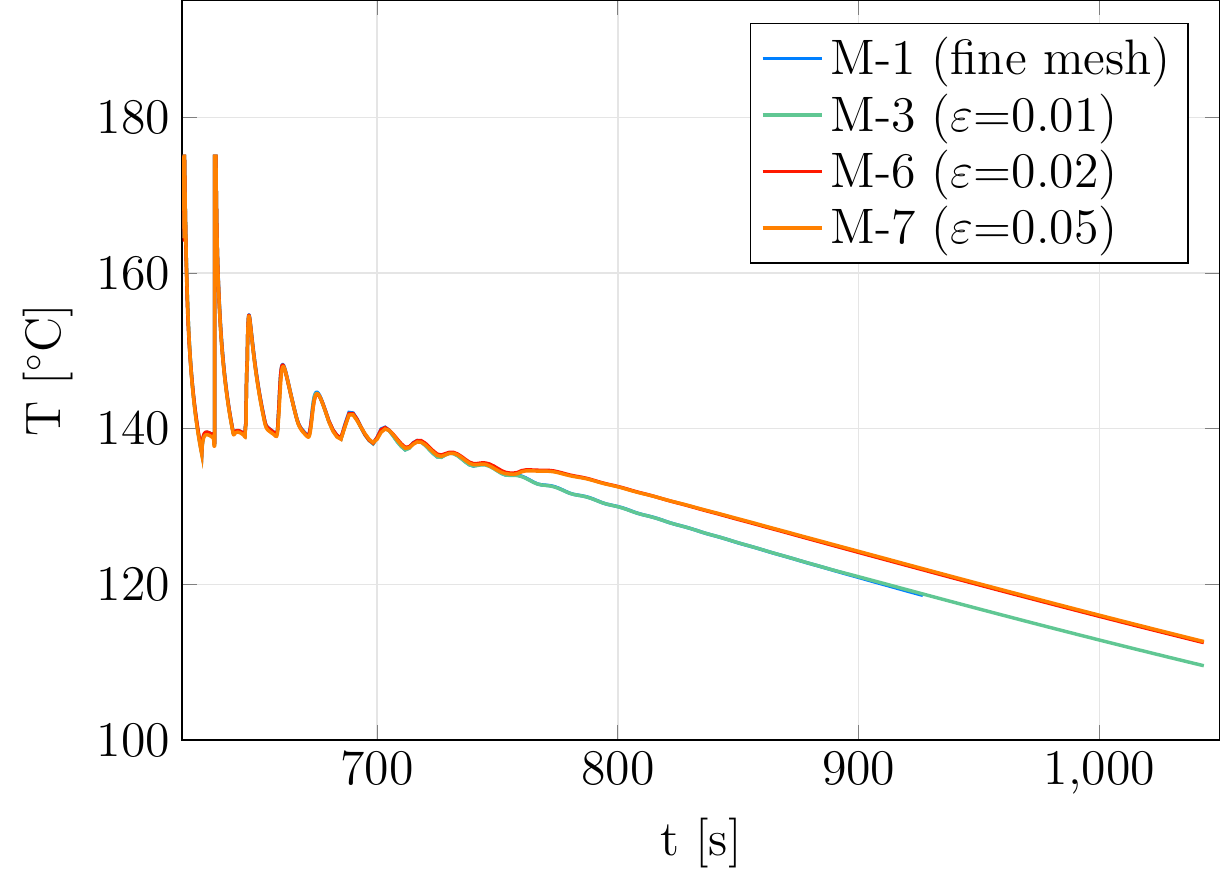}}
\caption{Influence of $\varepsilon$ }
\label{fig:Tt_eps}
\end{figure}

\begin{figure}[H]
\centering
\subfloat[M-3 ($\varepsilon$=0.01)]{\includegraphics[width=0.29\linewidth]{M3T175h_no79}}
\subfloat[M-6 ($\varepsilon$=0.02)]{\includegraphics[width=0.283\linewidth]{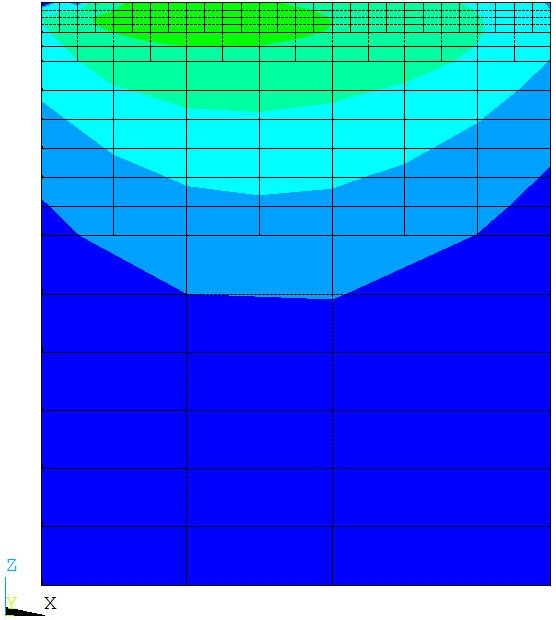}}
\subfloat[M-7 ($\varepsilon$=0.05)]{\includegraphics[width=0.283\linewidth]{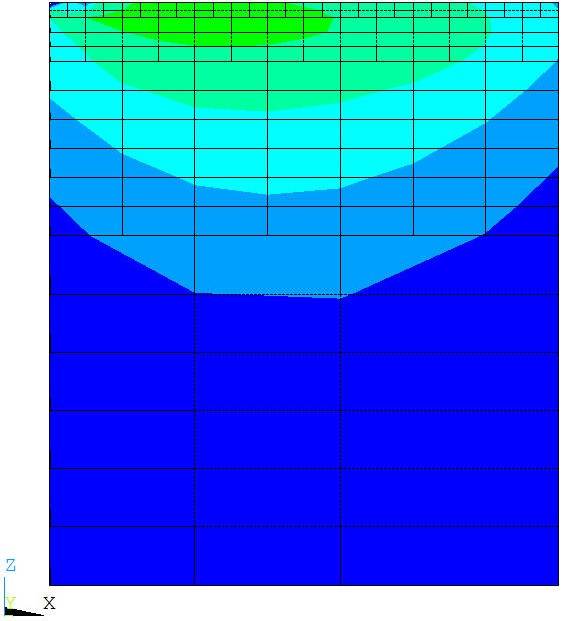}}
\subfloat{\includegraphics[scale=0.3]{legend_vert}}
\caption{Temperatures in last time step of the final coarsened meshes for models with varying $\varepsilon$. Cross-sectional plane at  y=$\frac{4}{7}\cdot$l}
\label{fig:contour_epsilon}
\end{figure}

\paragraph{Variation of numerical parameters}
The default remeshing and coarsening parameters as listed in table \ref{tab:def_param} were applied in simulations M-2 and M-3. Varying the parameters nh$_{\text{add}}$ and $\varepsilon$ can influence the efficiency and/or accuracy of the simulations compared to the default model M-1.\\
\indent The first parameter that was varied is nh$_{\text{add}}$. This parameter influences the computational time as it controls the trade-off between the number of inactive dofs present in remeshing step versus the number of times remeshing occurs. To investigate this, the simulations were additionally performed for nh$_{\text{add}}$=2 (\textbf{M-4}) and nh$_{\text{add}}$=4 (\textbf{M-5}). Table \ref{tab:def_param} shows that the computational time is similar for the models with one (M-3) and two quiet layers (M-4). However, the computational time was increased by 15\% when four quiet layers were added instead of one. This implies that it is more efficient to remesh often even if it means going through preprocessing more often vs. remeshing less frequently and having more number of dofs in each discretized geometry. The accuracy is also compared by looking at $T(t)$ of the aforementioned models (figure \ref{fig:Tt_nhadd}) and the temperature contour plots at the last time step of the last remeshing step (figure \ref{fig:contour_nhadd}). There is great agreement between the fine model and the coarsened models with varying nh$_{\text{add}}$.\\
\indent Next, the effect of coarsening parameter $\varepsilon$ was investigated. Since the simulation M-3 with $\varepsilon$=0.01 already yielded very accurate results compared to the results of the fine model, the value of $\varepsilon$ was varied between 0.01 and 0.05 for \textbf{M-6} and \textbf{M-7} (table \ref{tab:def_param}). The comparison of $T(t)$ between the fine model M-1 and the models with varying $\varepsilon$ is shown in figure \ref{fig:Tt_eps}. It can be seen that even for $\varepsilon$=0.05, there is still good agreement with the fine model, especially at the initial stages of the simulation. The heating and reheating peaks upon element activation were accurately captured. Small differences can be observed after the influence of the nozzle fades for $\varepsilon$=0.05 (M-7) as the temperatures measured in M-7 are approximately 5\% higher compared to M-1. This is an indicator that coarsening occured too early at a certain point in the simulation. However, the results at node 1 show that this difference decreases again as a steady-state is approached. The global temperatures are displayed for the last time step in the last remeshing step (figure \ref{fig:contour_epsilon}). The results are again very similar, but it can be seen that M-3 only reached two levels of coarsening whereas M-6 and M-7 both reached three levels of coarsening. This explains the difference in computational time, which is listed for each model in table \ref{tab:def_param}. For $\varepsilon$=0.05 the computational time was reduced by approximately 15\% compared to $\varepsilon$=0.01.

\subsection{Homogenized infill structures}
\label{subsec:valid_infill}
\begin{figure}
\centering
\subfloat[Geometry with a fine discretization]{\includegraphics[width=0.3\linewidth]{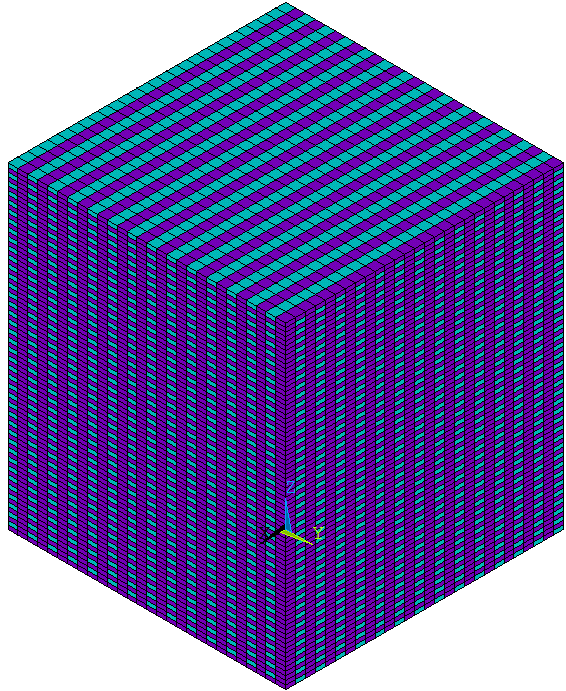}}\hspace{1em}
\subfloat[Geometry with a coarsened discretization]{\includegraphics[width=0.3\linewidth]{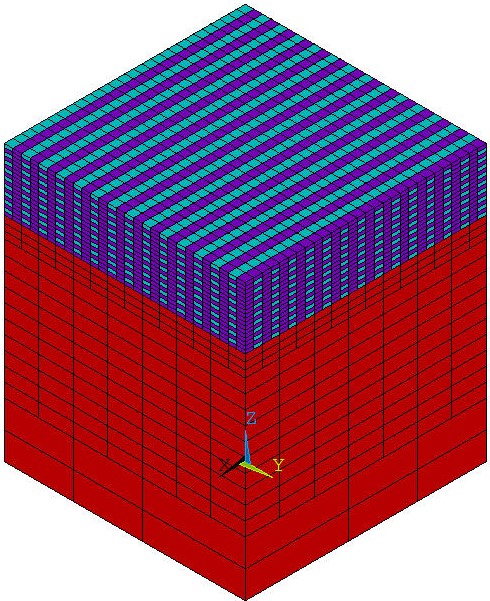}}
\caption{Mesh for infill geometries with a 50\% infill density in final remeshing step. Cyan: PLA elements, purple: air elements, red: homogenized elements}
\label{fig:mesh_infill}
\end{figure}

\begin{figure}
\centering
\subfloat{\includegraphics[width=0.3\linewidth]{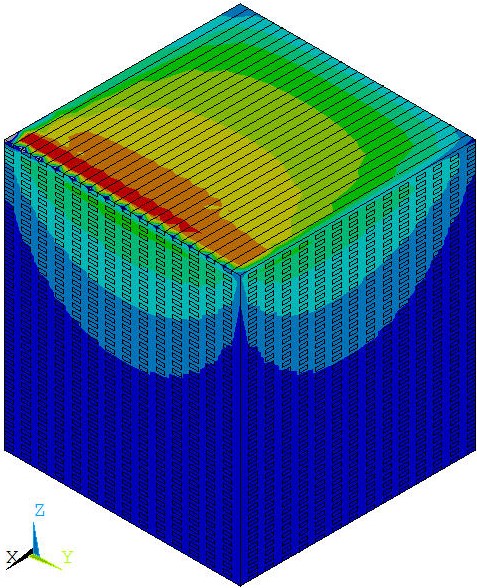}}\hspace{1em}
\subfloat{\includegraphics[width=0.3\linewidth]{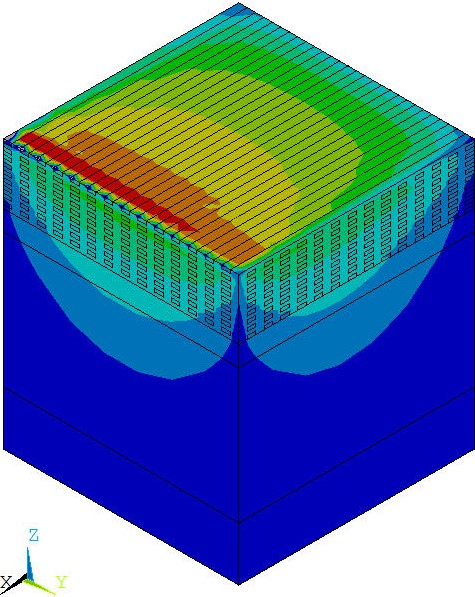}}\\
\subfloat{\includegraphics[width=0.315\linewidth]{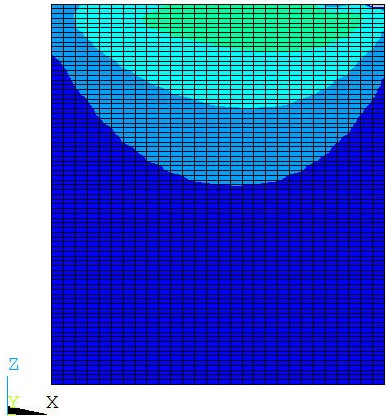}}\hspace{1em}
\subfloat{\includegraphics[width=0.3\linewidth]{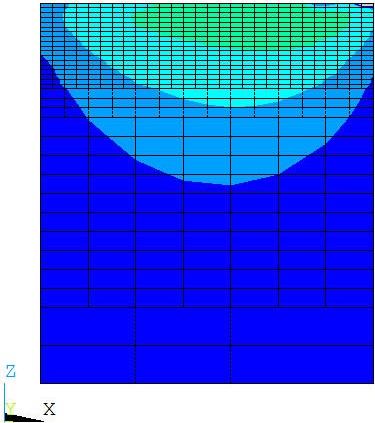}}\\
\subfloat{\includegraphics[scale=0.32]{legend_hor}}
\caption{Temperatures in the last time step of the final meshes with an infill geometry. Left: Fine mesh, right: coarsened mesh. Cross-sectional plane at  y=$\frac{4}{7}\cdot$l}
\label{fig:cont_infill}
\end{figure}

The applicability of the simulation framework on geometries with an air-filled infill pattern was tested in a similar fashion as was done in the previous section: the results of the thermal analysis on the geometry with a fine mesh was compared to the results of the thermal analysis on the same geometry but with a discretization that included adaptive coarsening. Remeshing was applied in both simulations. A rectilinear infill pattern with an infill density of 50\% was applied in both models. The model without any coarsening is described by model \textbf{M-8} whereas the model which includes coarsening is described by model \textbf{M-9}. All remeshing and coarsening parameters for these models are listed in table \ref{tab:def_param}.\\
\indent The meshes and temperature contour plots from the last remeshing step are shown in figure \ref{fig:mesh_infill} \& \ref{fig:cont_infill} respectively. Figure \ref{fig:cont_infill} shows that there is very good agreement between M-8 and M-9. The temperatures in the fine and coarsened parts of the mesh in model M-8 coincided very well with that of the fully fine mesh. It confirms that the homogenized material properties that were assigned to the coarse elements managed to capture the thermal behaviour of the air-PLA infill quite well.\\
\indent Model M-8 reached three levels of coarsening, even though the number of fine non-coarsened layers was relatively high (figure \ref{fig:mesh_infill}). Compared to the geometry with a dense infill (M-7 from figure \ref{fig:contour_epsilon}), there were more fine layers present in the coarsened model where the infill density is 50\%. It seems that the coarsening condition was satisfied more easily once the elements are homogenized, compared to the first coarsening level where the transition occured from the heterogeneous mesh to the homogenized mesh. When comparing the computational times, it can be seen that M-8 was solved significantly faster than its dense counterpart M-2, despite having the same total number of dofs. The difference can be attributed to the decrease in the total number of time steps for M-8. Even though the coarsened model M-9 was still more efficient than M-8, it lagged behind the efficiency of the densely coarsened model M-7. It seems that the significant increase in number of dofs outweighed the decrease in the total number of time steps. 

\subsection{Bridge Geometry}
\begin{figure}[H]
\centering
\subfloat[Bridge geometry]{\includegraphics[width=0.4\linewidth]{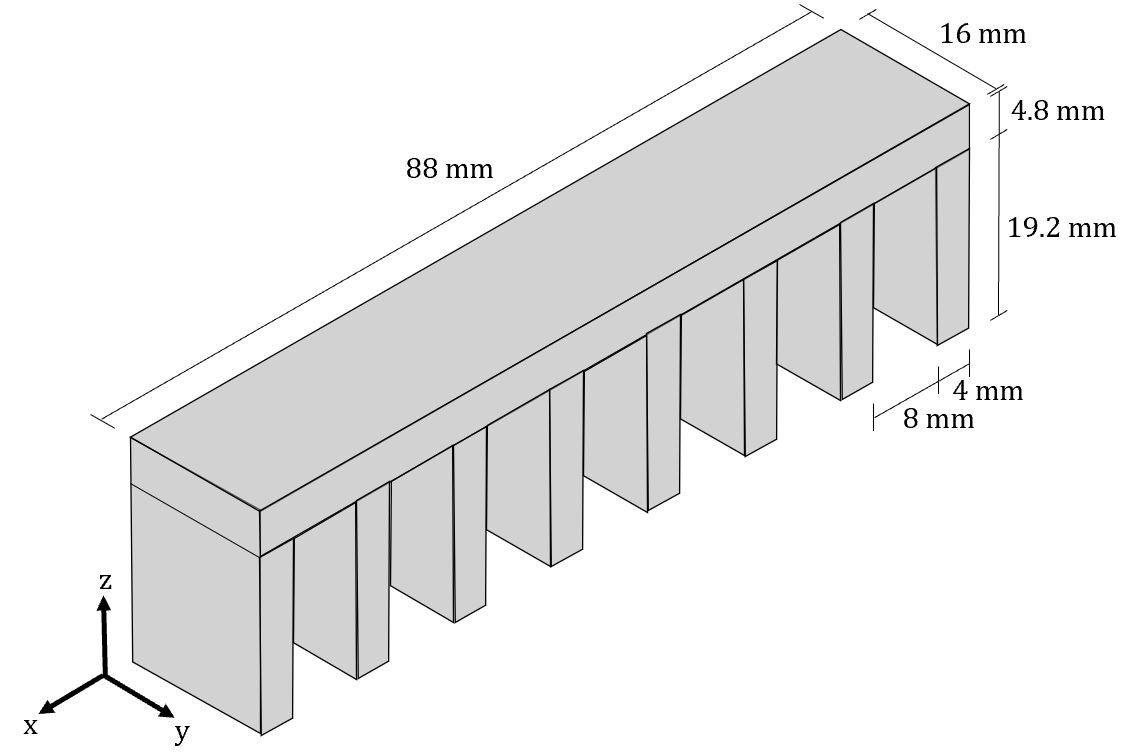}}
\subfloat[Final coarsened mesh (cyan: PLA, purple: air, red: homogenized elements)]{\includegraphics[width=0.33\linewidth]{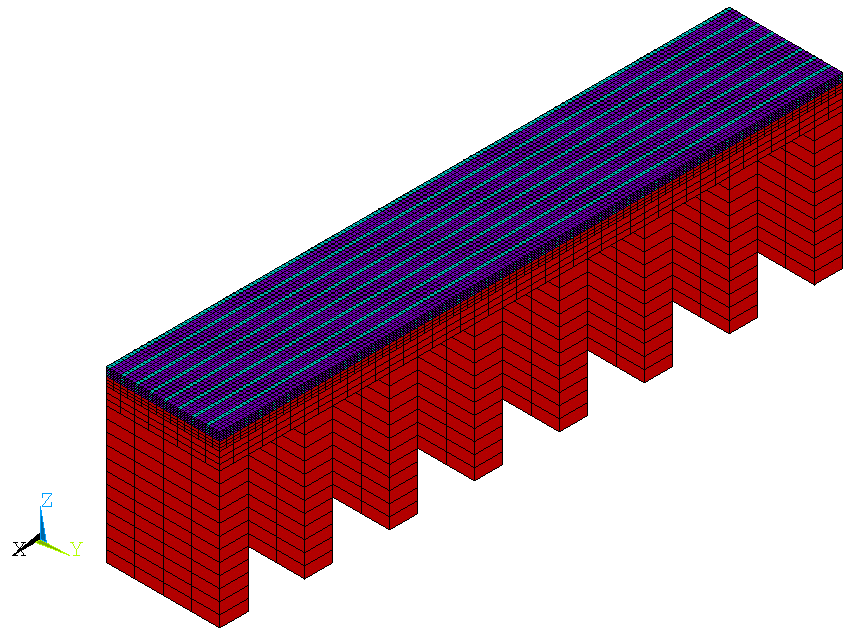}}\hspace{1em}
\subfloat[Infill pattern in the fine part of the mesh (25\% infill, only PLA shown)]{\includegraphics[width=0.2\linewidth]{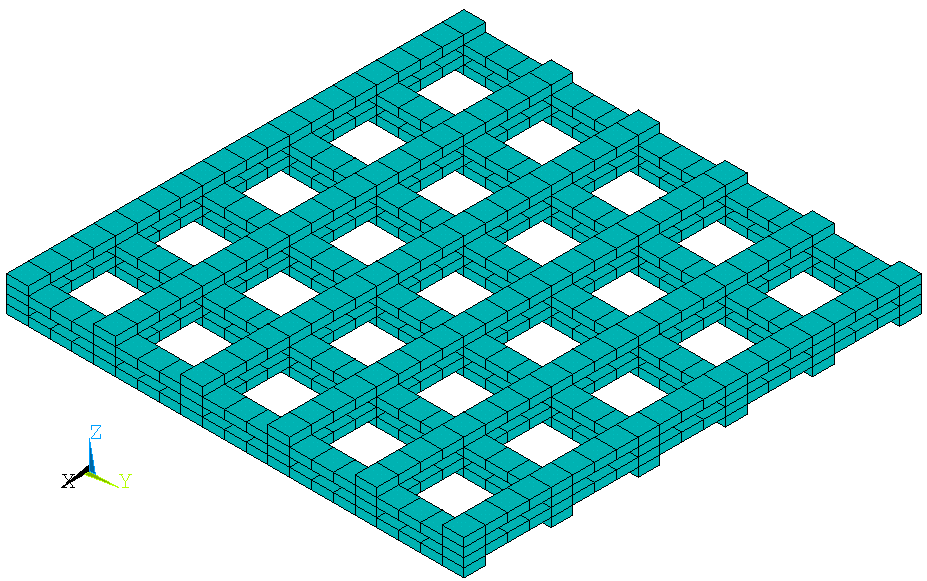}}
\caption{Bridge model}
\label{fig:bridge_mesh}
\end{figure}

\begin{figure}[H]
\centering
\subfloat[25\% of the print completed - 0$^{\circ}$ printing direction]{\includegraphics[width=0.45\linewidth]{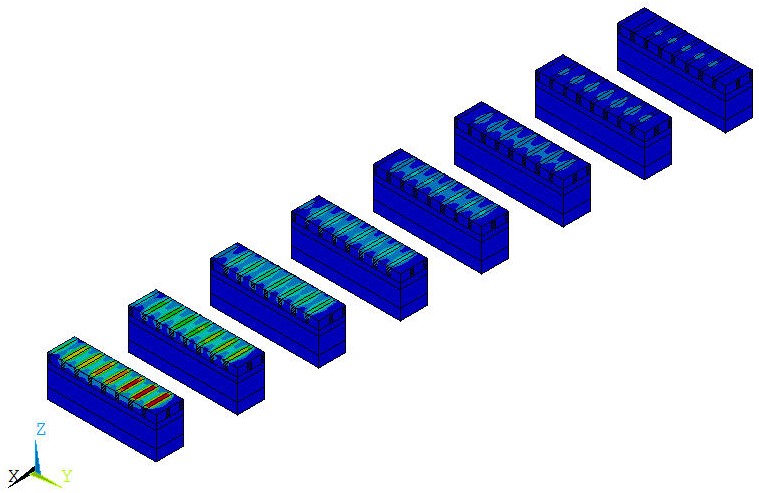}}\hspace{1em}
\subfloat[75\% of the print completed - 0$^{\circ}$ printing direction]{\includegraphics[width=0.45\linewidth]{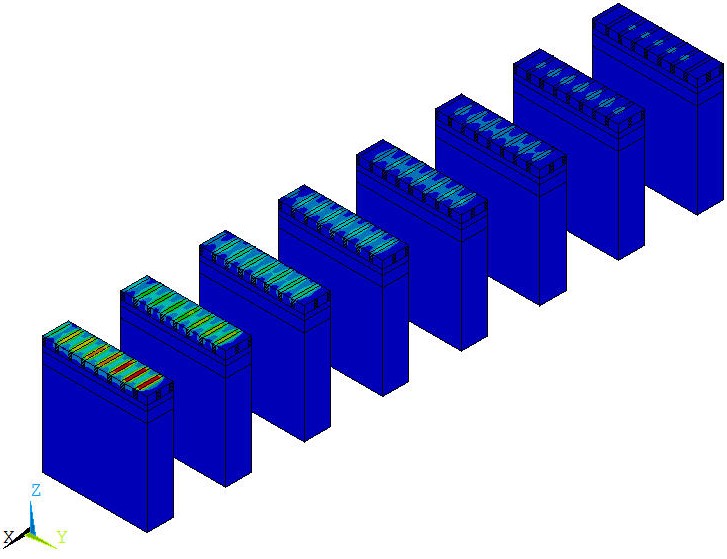}}\\
\subfloat[100\% of the print completed - 0$^{\circ}$ printing direction]{\includegraphics[width=0.45\linewidth]{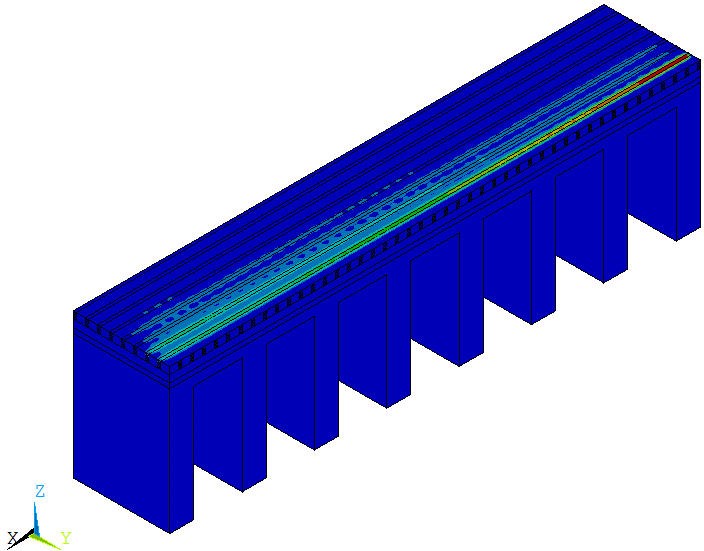}}\\
\subfloat{\includegraphics[scale=0.32]{legend_hor}}
\caption{Temperature contour plots during various printing stages of the bridge geometry}
\label{fig:bridge_temp}
\end{figure}

All of the numerical methods presented in the previous sections were tested on a bridge geometry (figure \ref{fig:bridge_mesh}). This geometry is inspired by \cite{Cattenone2019} and a rectilinear infill pattern with an infill percentage of 25\% was applied (figure \ref{fig:bridge_mesh}). The homogenized material properties assigned to the coarsened elements were calculated with equations \ref{eq:effect_cond}-\ref{eq:alpha}. All process parameters which were used to simulate the printing process are listed in table \ref{tab:proc_param}. A time step size 0.267s was applied. The initial and boundary conditions were applied in a similar fashion as for the block geometry (section \ref{subsec:loads_bcs}). The main difference is that the bridge geometry has additional external free surfaces which were subjected to convective heat transfer. The coarsening parameters were chosen as follows:
\begin{itemize}
\item MLVL=3
\item CF=2
\item nh$_{\text{add}}$=2 
\item $\varepsilon$=0.05
\end{itemize}
\indent The final coarsened mesh from the last remeshing step (figure \ref{fig:bridge_mesh}) shows that three levels of coarsening were reached in the majority of the geometry. There are less fine layers present compared to the block geometry (figure \ref{fig:mesh_infill}) which could be attributed to the larger overall printing time of the bridge, more cooling time for each printed layer and thus easier satisfaction of the coarsening condition. The maximum number of coarsening levels was restricted by the width of the bridge pillars, but the significant amount of layers in the coarsest level would otherwise allow for additional coarsening levels.\\
\indent The temperature contour plots are shown in figure \ref{fig:bridge_temp} for various stages of the printing process. For each remeshing step shown, the temperatures are plotted as the PLA filaments are printed in the 0$^{\circ}$-direction (global x-direction). It can be seen that the coarsened layers have an even temperature distribution, whereas the influence of the printed filaments only reaches a few (fine) layers below the printed layer. This is an observation which is in agreement with the experimental measurements. The number of fine layers remains fairly constant throughout the various printing stages. Overall, the presented simulation framework also seems to work well on more complex geometries.   

\section{Conclusion}
\label{sec:conclusion}
A simulation framework has been presented for the transient thermal analysis of the fused filament fabrication (FFF) printing process. A hybrid element activation approach and adaptive coarsening have been applied to reduce the computational expense that is normally attached to heat transfer simulations of the FFF printing process. Additionally, the objective was to minimize the loss of accuracy and to have an accurate representation of the deposition physics.\\
\indent The simulation framework has been validated with experimental thermal measurements that were performed during the printing of a block. The comparison showed good agreement between the results, especially towards the center of the printed block. An important finding of this research is that the activation temperature in the simulations has to be significantly lower than the nozzle temperature. This is often not considered in heat transfer simulations of the FFF printing process.\\
\indent The applicability of the framework was also assessed by looking at its numerical accuracy and efficiency for geometries with both a dense infill and an air-filled infill pattern. The computational time of the simulations on meshes where both remeshing and adaptive coarsening were applied was approximately one fifth of the computational time simulations on a geometry with a fine mesh without any efficiency measures. Moreover, great agreement was found between the models, both locally and globally. Good agreement was also found between the models in which the part had an air-filled infill pattern. The homogenized material properties based on the infill density on the printed part which were assigned to the coarsened part of the mesh were able to capture the thermal behaviour of heterogeneous material well. \\ 
\indent Overall, the framework shows great potential for efficiently simulating heat transfer in the FFF printing process. The framework could be further improved by optimization of the time step size and a more accurate load prescription. The next step would be to extend the simulation framework to thermo-mechanical simulations for residual stress prediction. 

\clearpage
\bibliography{Coarsening_paper}
\clearpage
\noindent \textbf{Author contributions}: Conceptualization: Nathalie Ramos and Josef Kiendl, Methodology: Nathalie Ramos and Christoph Mittermeier, Validation: Nathalie Ramos, Writing - original draft: Nathalie Ramos, Writing - review \& editing: Josef Kiendl\\
\\
\textbf{Funding}: This research 'Efficient simulation of the heat transfer in fused filament fabrication' is funded by dtec.bw - Digitalization and Technology Research Center of the Bundeswehr. Dtec-bw is funded by the European Union - NextGenerationEU. This work was also supported by the European Research Council through the H2020 ERC Consolidator Grant 2019 n. 864482 FDM$^2$. The support of AMOS under project 3: Risk management and maximized operability of ships and ocean structures is also gratefully acknowledged.\\
\\
\textbf{Data availability}: The data presented in this study are available on request from the corresponding author.\\ 
\\
\textbf{Conflicts of interest}: The authors declare no conflict of interest. 

\end{document}